 \documentclass[ reprint,superscriptaddress,longbibliography,amsmath,amssymb,aps,pra,
]{revtex4-2}

\usepackage{graphicx}
\usepackage{dcolumn}
\usepackage{hyperref}
\usepackage[mathlines]{lineno}
\usepackage{balance}

\usepackage{physics}
\usepackage{microtype}
\usepackage{subcaption}
\usepackage{soul}
\usepackage{bbm}
\usepackage{bbold}
\usepackage{multirow}
\usepackage{array}
\usepackage{stackengine}
\usepackage{titlesec}

\usepackage{booktabs}
\usepackage{multirow}
\usepackage{array}

\usepackage{natbib}

\usepackage{float}

\usepackage{xcolor}
\usepackage{xspace}
\usepackage[normalem]{ulem}
\usepackage{csquotes}
\usepackage{placeins}

\definecolor{patriarch}{rgb}{0.5, 0.0, 0.5}
\definecolor{darkraspberry}{rgb}{0.53, 0.15, 0.34}
\definecolor{brinkpink}{rgb}{0.98, 0.38, 0.5}
\definecolor{skobeloff}{rgb}{0.0, 0.48, 0.45}
\definecolor{mypink}{RGB}{226,68,130}
\definecolor{darkpastelgreen}{rgb}{0.01, 0.75, 0.24}
\definecolor{pigmentgreen}{rgb}{0.0, 0.65, 0.31}

\usepackage{tikz}
\usetikzlibrary{
    positioning,   
    arrows.meta,   
    calc,          
    shapes.geometric 
}

\usepackage{bm}
\usepackage{comment}
\usepackage{hyperref}
\hypersetup{
    colorlinks=true,
    urlcolor=blue,
	citecolor=blue,
	allcolors=blue
}

\begin{document}

\preprint{APS/123-QED}

\title{Superposed quantum evolutions across chaotic and regular regimes} 
\author{Amit Anand}
\email{a63anand@uwaterloo.ca}
\affiliation{Institute for Quantum Computing, University of Waterloo, Waterloo, Ontario, Canada N2L 3G1}
\affiliation{Department of Physics and Astronomy, University of Waterloo, Waterloo, Ontario, Canada N2L 3G1}
\author{Anne-Catherine de la Hamette}
\affiliation{University of Vienna, Faculty of Physics, Vienna Doctoral School in Physics, and Vienna Center for Quantum Science and Technology (VCQ), Boltzmanngasse 5, A-1090 Vienna, Austria}
\affiliation{Institute for Quantum Optics and Quantum Information (IQOQI),
Austrian Academy of Sciences, Boltzmanngasse 3, A-1090 Vienna, Austria}
\affiliation{Institute for Theoretical Studies, ETH Zürich, 8006 Zürich, Switzerland}
\author{Robert Mann}
\affiliation{Department of Physics and Astronomy, University of Waterloo, Waterloo, Ontario, Canada N2L 3G1}
\affiliation{Perimeter Institute for Theoretical Physics, 31 Caroline St N, Waterloo, Ontario, Canada N2L 2Y5}
\author{Shohini Ghose}
\affiliation{Institute for Quantum Computing, University of Waterloo, Waterloo, Ontario, Canada N2L 3G1}
\affiliation{Perimeter Institute for Theoretical Physics, 31 Caroline St N, Waterloo, Ontario, Canada N2L 2Y5}
\affiliation{Department of Physics and Computer Science, Wilfrid Laurier University, Waterloo, Ontario, Canada N2L 3C5}

\date{\today}

\begin{abstract}
While the superposition of quantum evolutions is known to produce interference effects, the interference between evolutions with regular and chaotic classical limits remains largely unexplored. Here, we use a Mach–Zehnder interferometer to investigate the superposition of two quantum evolutions, implemented via post-selection, and to compare it with the corresponding classical mixture. The quantum kicked top provides a natural platform for this study, as its classical dynamics ranges from regular to mixed to fully chaotic depending on the Hamiltonian parameters. We show that when a regular evolution is superposed with a chaotic one, the resulting subsystem entropy can exceed that of the classical mixture, provided the contribution of the chaotic branch dominates in the superposed quantum evolution. We further demonstrate that entropy production in such superpositions is strongly influenced by the structure of the underlying classical phase space. We further show that increased entropy generation can occur for purely regular dynamics at small values of the chaos parameter, given an appropriate choice of post-selection. These results reveal a nontrivial interplay between classical chaos and quantum interference in superposed quantum dynamics.
\end{abstract}

\keywords{Suggested keywords}
\maketitle


\section{Introduction}\label{sec:Introduction}

The principle of superposition lies at the heart of quantum mechanics and plays a central role in how quantum systems evolve. When two or more quantum evolutions are coherently combined, their complex amplitudes interfere, giving rise to characteristic quantum interference patterns. In this sense, the superposition of quantum evolutions is directly related to the familiar interference of waves: both arise from adding complex amplitudes and observing the resulting pattern. While superposition has been extensively explored across a wide range of physical settings, previous work has primarily focused on understanding how phase differences between multiple paths lead to constructive or destructive interference \cite{ou_MZI_1996, MZI_dowling_2000}. Such interference effects form the basis of many quantum–enhanced metrological protocols, where exploiting coherence allows measurements that surpass the classical shot-noise limit and approach the Heisenberg limit \cite{sanders_MZI_1995,MZI_HL_gerry_2000}. 

Surprisingly, however, the superposition of quantum evolutions whose classical limits exhibit different dynamical behaviour, such as regular versus chaotic dynamics, has not been systematically explored. Understanding how  a regular and a  chaotic evolution interfere presents a fundamentally new question, with potential implications for quantum information scrambling, quantum control, and our broader understanding of quantum-classical correspondence. In particular, it raises the question: \emph{How does classical chaos influence the interference pattern resulting from the superposition of two quantum evolutions?}

From a broader perspective, our setup is also motivated by ideas from the field of quantum reference frames, where changes of quantum frame can lead to superpositions of conditional dynamics when the reference system itself is in a quantum superposition (see e.g.,~\cite{Giacomini_2017_covariance, Vanrietvelde_2018a, delaHamette_2021_indefinitemetric}). In that context, relational descriptions often give rise to coherently controlled evolutions of subsystems. While the interferometric scheme studied here does not implement a quantum reference frame transformation in the technical sense, it realizes an analogous structure in which distinct dynamical evolutions are coherently combined and post-selected. Our results can therefore be viewed as probing how such superposed dynamics is manifest in systems whose classical limits exhibit regular or chaotic behaviour.

Interferometry provides a powerful tool set for studying how quantum amplitudes combine. By splitting, evolving, and recombining wave packets, interferometric techniques extract information from the resulting interference pattern. Interferometers are engineered for a wide range of tasks -- including gravitational-wave detection~\cite{gravity_sensing_2011}, Ramsey spectroscopy~\cite{giovannetti_metrology_2011}, and atomic or optical sensing -- typically with the goal of estimating a phase parameter with maximal precision. The Mach-Zehnder interferometer (MZI) is the paradigmatic example, widely used in imaging, sensing, and information processing. When supplied with non-classical probe states such as NOON states~\cite{sanders_MZI_1995,MZI_HL_gerry_2000,MZI_dowling_2000,MZI_campos_2002}, the MZI can attain Heisenberg-limited scaling, spurring extensive work on quantum metrology protocols~\cite{ou_MZI_1996,MZI_phase_bouwmeester_2007,Schmiedmayer_MZI_2009}.

In this work, we repurpose the MZI architecture to study an entirely different question: the interference between two quantum kicked top evolutions characterized by different values of the chaos parameter $\kappa$. To our knowledge, this is the first investigation of superposing quantum evolutions that correspond, in the classical limit, to dynamically distinct behaviours. Instead of inserting a relative phase shift between the arms, each arm of the interferometer contains a kicked top unitary with a different parameter. The contribution of each evolution to the final interferometric state is controlled by the mixing angles of the beam splitters. The quantum kicked top is an ideal test bed for such an exploration: it is a finite-dimensional spin-$j$ Floquet system with a well-understood classical limit whose degree of chaos can be tuned continuously \cite{haake1987}. It has been extensively used to study quantum chaos \cite{Ghose_Sanders_ent_dynamics_2004,wang_ghose_entg_2004,chaudhury_2009,Lombardi_ent_2011,anand_2021_simulating}, entanglement generation \cite{Ghose_chaos_ent_dec_2008,Madhok_Dogra_Lakshminarayan_2018,untangling_ghose_meenu_2019}, quantum-classical correspondence \cite{haake1987,Kumari_Ghose_vicinity_2018}, and quantum recurrences \cite{anand_davis_ghose_2024}.

By evolving an initial spin coherent state, tensored with an ancilla qubit, through two kicked top unitaries in the MZI, we uncover several striking phenomena upon recombination. The respective values of the chaos parameter $\kappa$, which determines whether the classical dynamics is regular, mixed, or chaotic, strongly influences how the two evolutions interfere. In particular, we show that the time-averaged entropy of the spin subsystem can increase or decrease depending on whether the two evolutions correspond to similar or different dynamical regimes. When one evolution is classically regular and the other chaotic, the superposition typically leads to \emph{greater} entropy generation than the classical mixture of both evolutions, provided the contribution of the chaotic path dominates. Conversely, superposing two regular or two chaotic evolutions does not increase subsystem entropy relative to the classical mixture.  We further demonstrate that the mixing angles of the beam splitters play a crucial role: appropriate post-selection can amplify entropy production even when the chaotic path initially has smaller weight, connecting it to well-known \textit{quantum eraser} phenomena \cite{scully_quantum_erasure_1982,scully_quantum_erasure_2000,monken_quantum_erasure_2002}. 
Overall, our results reveal a rich and nontrivial interplay between classical dynamical behaviour and quantum interference. They indicate that superposing distinct quantum evolutions can reveal signatures associated with regular and chaotic behaviour, even in finite-dimensional systems away from the semiclassical limit.


\section{Background}

The quantum kicked top (QKT) is a finite-dimensional dynamical model used to study quantum chaos, known for its compact phase space and parameterizable chaoticity structure \cite{haake1987}.  The time-dependent, periodically driven system is governed by the Hamiltonian,
\begin{equation}\label{eq:Floquet_unitary_kappa}
 H = \hbar\frac{\alpha J_{y}}{\tau} +  \hbar \frac{\kappa J_{z}^2}{2j} \sum_{n=-\infty}^{\infty} \delta (t-n\tau),
\end{equation}
where $\{ J_{x}, J_{y}, J_{z}\}$ are the generators of angular momentum: $[J_i, J_k] = i\epsilon_{ikl} J_l$.  It describes a spin of size $j$ precessing about the $y$-axis together with impulsive state-dependent twists about the $z$-axis with magnitude characterized by the chaoticity parameter $\kappa$.  The period between kicks is $\tau$, and $\alpha$ is the amount of $y$-precession within one period. The quantum kicked top Hamiltonian commutes with the total angular momentum operator $J^2$, $[H,J^2] = 0$. Hence, it can be considered as an $N = 2j$ qubit system \cite{kumari_2019_quantumclassical} confined to the symmetric subspace of $(\mathbb C^2)^{\otimes N}$. The spin-$j $ operators are written in terms of the single qubit Pauli rotation operators as  
\begin{equation}
J_{\alpha}=\frac{1}{2} \sum_{i=1}^{2 j} \sigma_{i \alpha}, \quad \alpha \in\{x, y, z\}
\end{equation}
where $\sigma_{i\alpha}$ denotes $\sigma_{\alpha}$ acting on the $i^{th}$ qubit. The qubit version of the quantum kicked top enables us to discuss subsystem entropy, a quantity of interest in the following sections of this paper. The associated Floquet time evolution operator for one period is
\begin{equation}\label{eq:floquet-unitary}
    U_\kappa = \exp\Big(-i\frac{\kappa }{2j}J_{z}^2\Big) \exp\Big(-i\alpha J_{y}\Big)
\end{equation}
where $\kappa$ indexes the degree of chaoticity of the classical dynamics.

\begin{figure*}[t]
\centering

\begin{subfigure}[t]{0.48\textwidth}
    \centering
    \includegraphics[width=\textwidth]{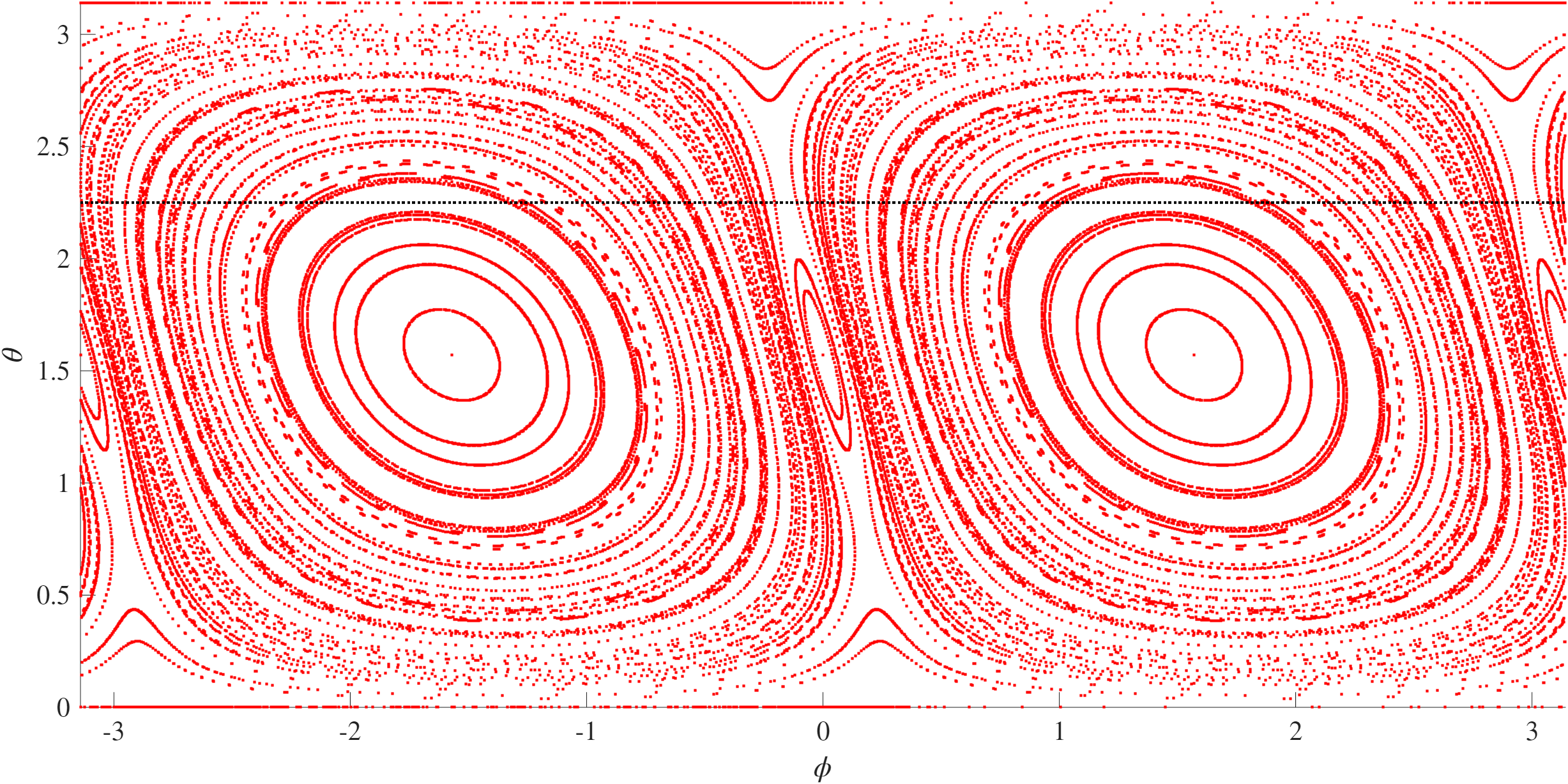}
    \caption{$\kappa = 0.5$}
    \label{fig:classical_0.5}
\end{subfigure}
\hfill
\begin{subfigure}[t]{0.48\textwidth}
    \centering
    \includegraphics[width=\textwidth]{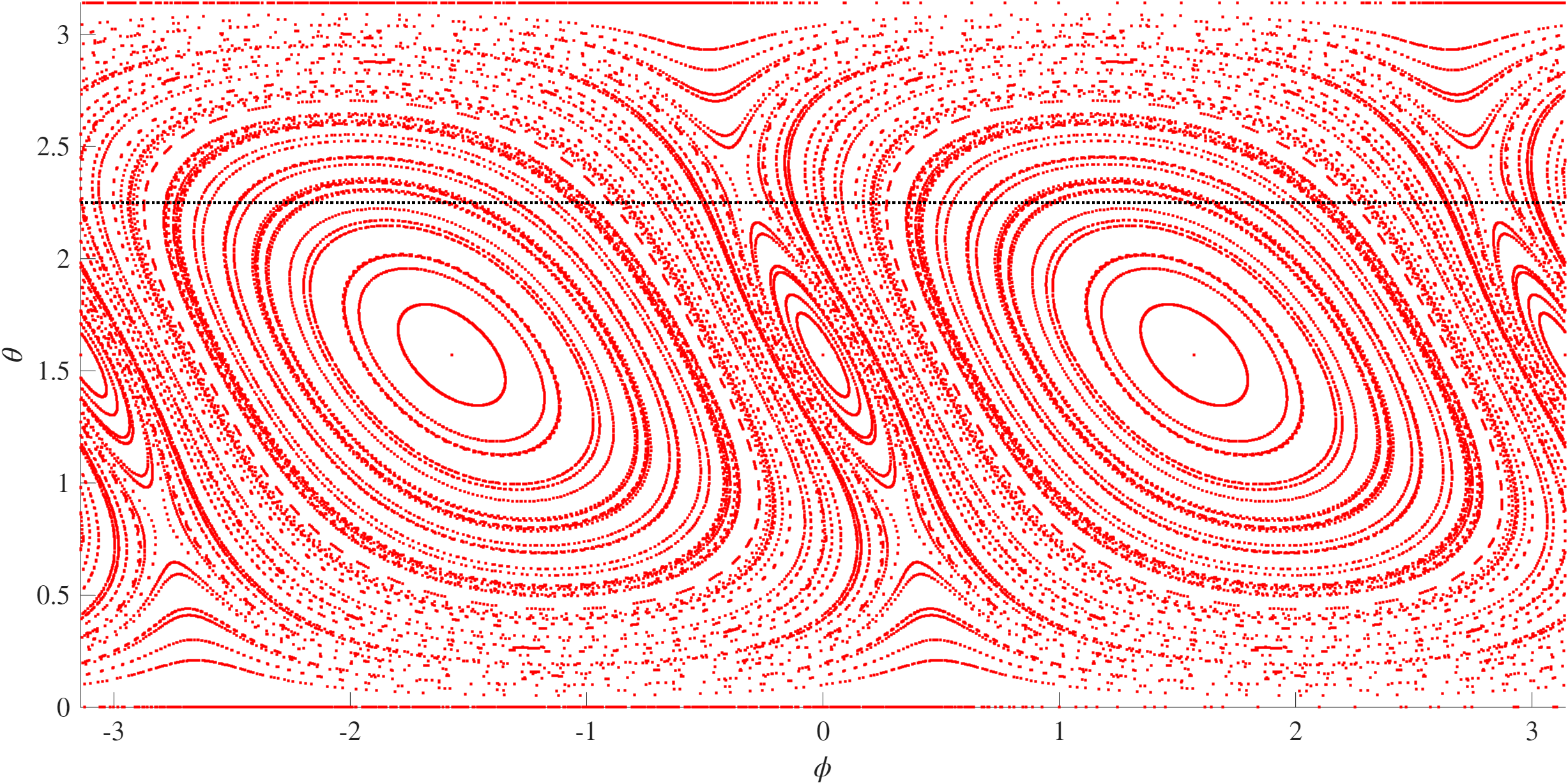}
    \caption{$\kappa = 1$}
    \label{fig:classical_1}
\end{subfigure}

\vspace{0.2cm}

\begin{subfigure}[t]{0.48\textwidth}
    \centering
    \includegraphics[width=\textwidth]{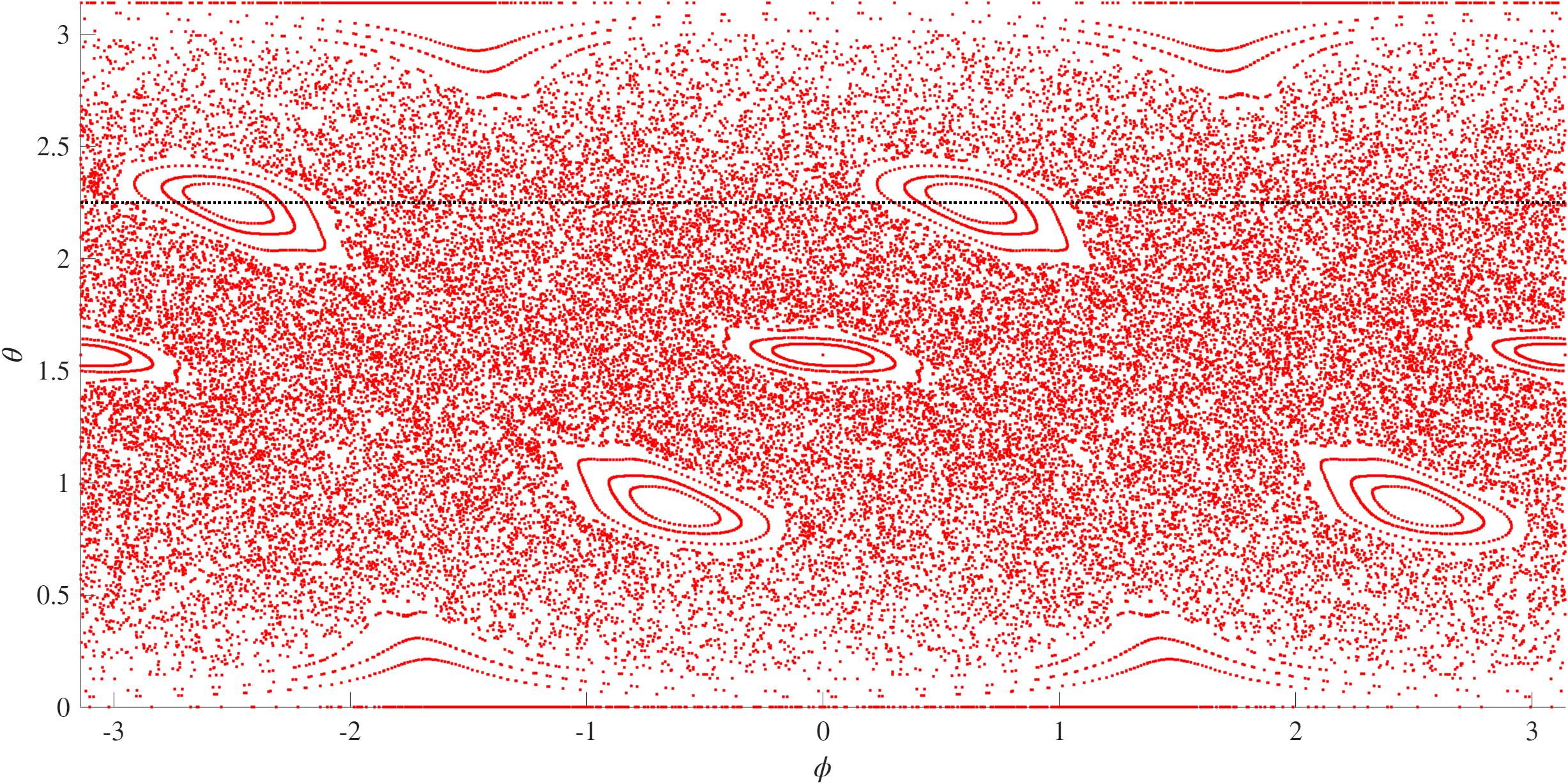}
    \caption{$\kappa = 3$}
    \label{fig:classical_3}
\end{subfigure}
\hfill
\begin{subfigure}[t]{0.48\textwidth}
    \centering
    \includegraphics[width=\textwidth]{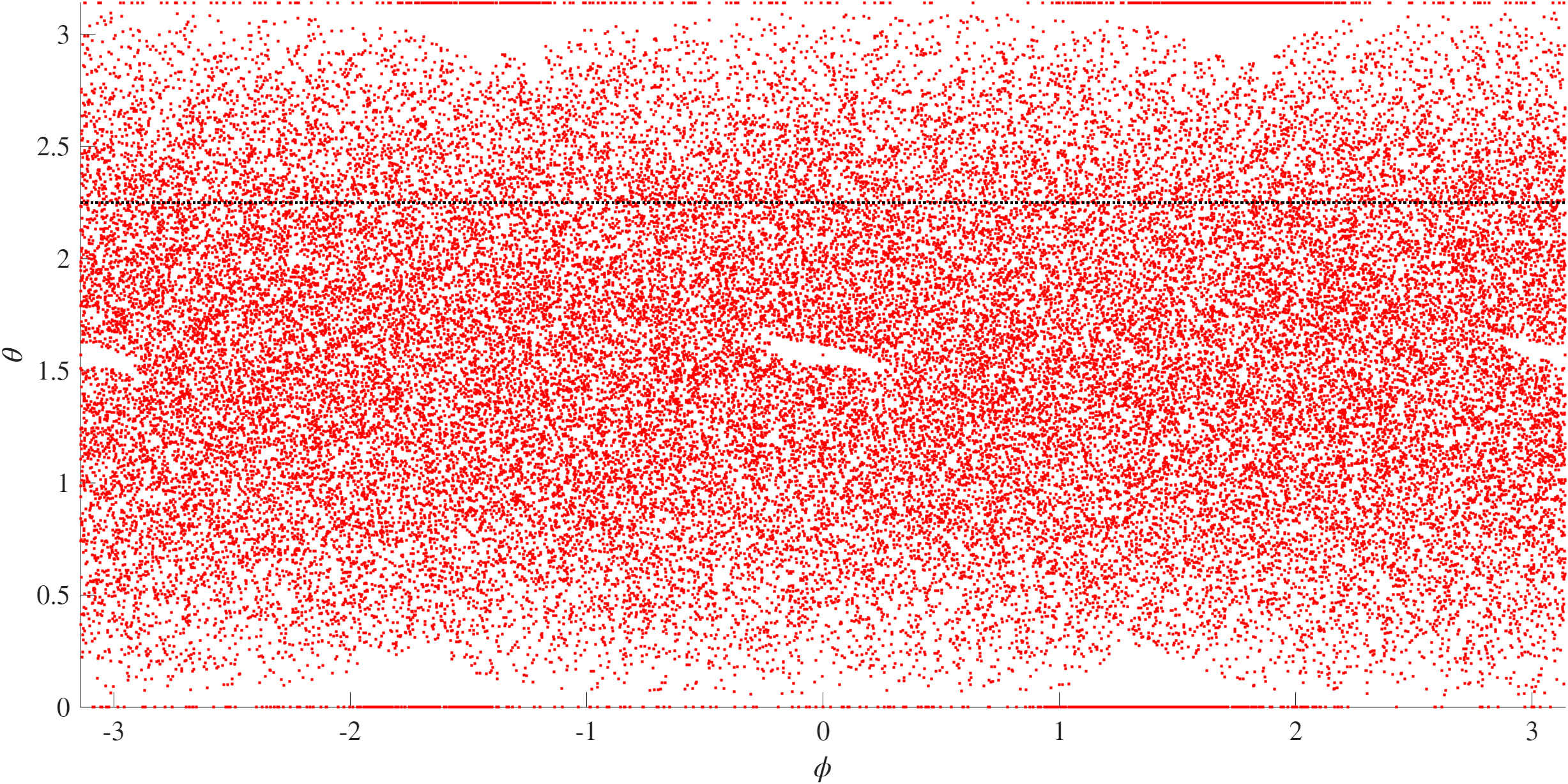}
    \caption{$\kappa = 6$}
    \label{fig:classical_6}
\end{subfigure}

\caption{
Stroboscopic map showing the classical time evolution over 200 kicks for 289 initial states. Each initial state is evolved using the classical map in Eq.~\eqref{eq:classical_eom} for different values of  {the chaos parameter} $\kappa$ and  {the amount of $y$-precession per period} $\alpha=\pi/2$. The black dotted line indicates $\theta = 2.25$. As $\kappa$ increases, the dynamics of the initial states along the black dotted line evolve from regular motion to a mixed regular–chaotic regime, and ultimately to fully chaotic behaviour.
}
\label{fig:classical_map}
\end{figure*}

The classical kicked top can be obtained by computing the Heisenberg equations for the re-scaled angular momentum generators, $X_{i}=J_i/j$, satisfying $[X_i, X_k] = (1/j)\epsilon_{ikl} X_l$ and followed by the limit $j \to \infty$ \cite{haake1987}. Denoting the previous $X_i$ as $X$, $Y$, and $Z$, and for the commonly considered case of $(\tau=1, \alpha = \pi/2)$, we have the following recursion relations:
    \begin{align}\label{eq:classical_eom}
     X_{n+1} &= Z_n \cos(\kappa X_n)+Y_n \sin(\kappa X_n), \nonumber \\ 
     Y_{n+1} &= Y_n \cos(\kappa X_n)-Z_n \sin(\kappa X_n), \nonumber \\
     Z_{n+1} &= -X_n.
    \end{align}
As the chaoticity parameter $\kappa$ is varied, the classical dynamics ranges from completely regular motion ($\kappa$ $\leq $ 2.1) to a mixture of regular and chaotic motion (2.1 $\leq \kappa \leq 4.4$) to fully chaotic motion ($\kappa > 4.4$) \cite{Kumari_Ghose_vicinity_2018}. The classical stroboscopic map in polar coordinates for a set of initial conditions with $\kappa= 0.5$, $\kappa= 1.0$, $\kappa= 3.0$ and $\kappa=6.0$ is given in Figs.~\ref{fig:classical_0.5} -- \ref{fig:classical_6}, respectively.

In this investigation we  set $\alpha = \pi/2$ without  loss of generality. For $\kappa = 0.5$ and $\kappa=1$, the phase space is fairly regular as shown in Figs.~\ref{fig:classical_0.5} and~\ref{fig:classical_1}. As  $\kappa$ gets larger the chaotic sea increasingly dominates;  regular islands get localized and finally vanish for sufficiently high values of $\kappa$.  When comparing the quantum and classical systems, a common approach is to quantify the similarity between the quantum dynamics of a quantum state (usually spin coherent) with the classical dynamics of a classical state (usually a point with the same centroid as the spin coherent state). The spin coherent state (SCS) $\ket{\theta,\phi}$ is defined as a rotated version of the highest-weight Dicke state~\cite{arecchi_1972}:
\begin{equation}\label{eq:spin coherent state}
    \ket{\theta,\phi} = e^{-i\phi J_z}\, e^{-i\theta J_y}\ket{j,j}.
\end{equation}
SCS states \cite{radcliffe_1971_some} are  minimum uncertainty states in spin systems  that saturate the uncertainty relation 
\begin{equation}
    \Delta J_i\Delta J_j = \frac{\hbar}{2}|\Delta J_k|,
\end{equation}
where $i$, $j$ and $k$ are permutations of $x,y$ and $z$. The uncertainty for these states is distributed symmetrically over the two operators.


\section{Quantum controlled chaos parameters}

\begin{figure*}[htbp]
    \centering
    \begin{tikzpicture}[
        node distance=4cm and 4cm, 
        bs1/.style = {rectangle, draw=blue!50!black, fill=blue!10, very thick,
                     minimum size=1cm, label={[blue!70]90:BS1}},
        bs2/.style = {rectangle, draw=blue!50!black, fill=blue!10, very thick,
                     minimum size=1cm, label={[blue!70]90:BS2}},
        mirror/.style = {rectangle, draw=gray!60!black, fill=gray!20, very thick,
                         minimum size=1cm, label={[gray!80]90:Mirror}},
        detector/.style = {trapezium, trapezium left angle=70, trapezium right angle=110,
                           draw=black, fill=gray!70, minimum width=1cm, inner sep=3pt,
                           label={[black]90:Detector}},
        beam/.style = {draw=red, very thick, -{Stealth[length=3mm, width=2mm]}},
        path_1/.style = {draw=cyan!70!black, thick, dashed, -{Stealth[length=3mm, width=2mm]}},
        path_2/.style = {draw=orange!70!black, thick, densely dotted, -{Stealth[length=3mm, width=2mm]}},
        beam_out/.style = {draw=purple!80!black, very thick, -{Stealth[length=3mm, width=2mm]}},
        evolution/.style = {rectangle, rounded corners, draw=green!50!black, fill=green!10,
                            thick, minimum width=2cm, minimum height=0.75cm},
        phaseshift/.style = {rectangle, rounded corners, draw=purple!50!black, fill=purple!10,
                             thick, minimum width=2cm, minimum height=0.75cm}
    ]
    
    \node (source) at (-4, 0) {\textbf{Source}};
    \node[bs1] (BS1) at (0,0) {$\sigma_1, \varphi_1$};
    \node[mirror] (M1) at (0,6) {M$_1$};
    \node[mirror] (M2) at (6,0) {M$_2$};
    \node[bs2] (BS2) at (6,6) {$\sigma_2,\varphi_2$};
    \node[detector] (D1) at (6,10) {D$_1$};
    \node[detector] (D2) at (10,6) {D$_2$};
    
    \draw[beam] (source) -- node[midway, below, yshift=-2mm, text=red] {Input: $\rho_{\kappa} \otimes \rho_{s}$} (BS1);
    
    \draw[path_1] (BS1) -- node[midway, left=3mm, text=cyan!70!black] {Path 1} (M1);
    \draw[path_1] (M1) -- (BS2);
    
    \draw[path_2] (BS1) -- node[midway, below=3mm, text=orange!70!black] {Path 2} (M2);
    \draw[path_2] (M2) -- (BS2);
    
    \draw[beam_out] (BS2) -- node[midway, right=3mm] {} (D1);
    \draw[beam_out] (BS2) -- node[midway, above=3mm] {} (D2);
    
    \node[evolution] at ($(BS1)!0.6!(M1)$) {Evolution $U^n_{\kappa_1}$};
    \node[phaseshift] at ($(BS1)!0.5!(M2)$) {Evolution $U^n_{\kappa_2}$};
    
    \end{tikzpicture}
    
    \caption{A Mach-Zehnder interferometer setup. The input system consists of a composite system of spin and ancilla qubit. The parameters $\sigma_1$ and $\sigma_2$ represent the beam splitter angles for $BS_1$ and $BS_2$, respectively. Path 1 consists of the $n$-fold application of the quantum kicked top unitary with  {chaos parameter} $\kappa_1$ ($U^n_{\kappa_1}$) while path 2 consists of the $n$-fold quantum kicked top unitary with  {chaos parameter} $\kappa_2$ ($U^n_{\kappa_2}$). The two detectors at the output ports are labeled $D_1$ and $D_2$.}
    \label{fig:mz_setup}
\end{figure*}
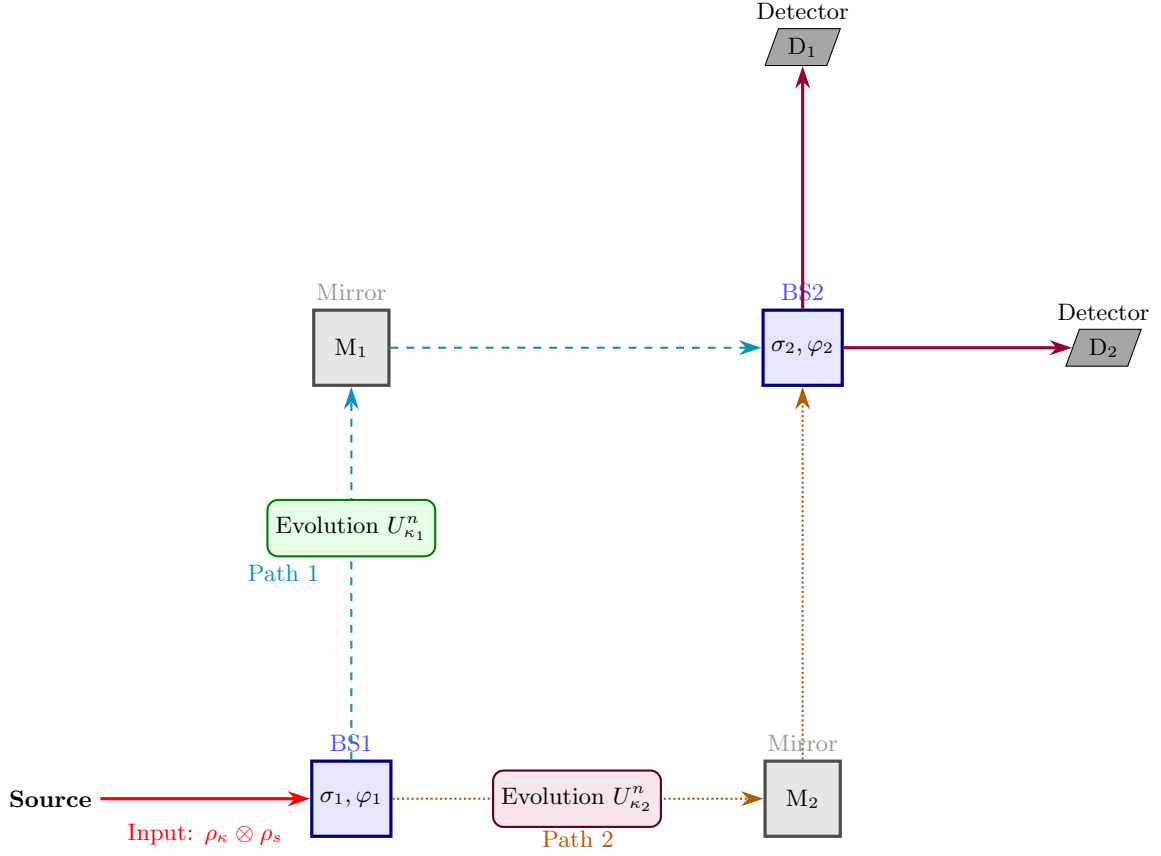
In this section, we introduce a modified version of the QKT model for studying superposed quantum evolution. We first promote $\kappa$ from a parameter to a Hermitian operator $\hat{\kappa}$. We modify the kicked top Hamiltonian as
\begin{equation}\label{eq:modified Hamiltonian 1}
 H = \hbar\alpha\,\frac{ I \otimes \hat{J}_{y}}{\tau}
   + \hbar\,\frac{\hat{\kappa} \otimes \hat{J}_{z}^{2}}{2j}
     \sum_{n=-\infty}^{\infty} \delta(t - n\tau),
\end{equation}
where $\hat{\kappa}\ket{\kappa_{1,2}} = \kappa_{1,2}\ket{\kappa_{1,2}}$, with $\kappa_{1,2}$ real eigenvalues and $\ket{\kappa_{1,2}}$ orthonormal states  spanning a Hilbert space isomorphic to $\mathbb{C}^2$.
We derive the unitary time evolution operator for one time period ($\tau$), given by
\begin{equation}\label{eq:floquet-unitary 1}
    U = \exp\!\Big(-i\,\frac{\hat{\kappa} \otimes \hat{J}_{z}^{2}}{2j}\Big)
        \exp\!\Big(-i\alpha\,(\hat{I} \otimes \hat{J}_{y})\Big),
\end{equation}
which can be written as a controlled operation on the spin system, controlled  on the $\kappa$-state, as
\begin{align}\label{eq:controlled unitary}
U = \ket{1}\bra{1} \otimes e^{-i\frac{\kappa_1}{2j}\hat{J}_{z}^{2}} e^{-i\alpha\hat{J}_{y}}
  + \ket{2}\bra{2} \otimes e^{-i\frac{\kappa_2}{2j}\hat{J}_{z}^{2}} e^{-i\alpha\hat{J}_{y}}
\end{align}
where we respectively identify the eigenstates $\ket{\kappa_{1,2}}$ as $\ket{1}$ and $\ket{2}$, corresponding to eigenvalues $\kappa_{1}$ and $\kappa_{2}$.

\subsection{Quantum kicked top in Mach-Zehnder interferometer}

A Mach–Zehnder interferometer is a phase-sensitive optical circuit that is widely used to study the interference pattern between two evolutions when they are superposed \cite{MZI_phase_bouwmeester_2007,MZI_phase_daniel_2013}. It consists of two beam splitters (BS) and two mirrors. In our setup, we use it to study the quantum controlled evolution of the quantum kicked top with different Hamiltonian parameters, and to analyze the effect of superposing two evolutions.

For our setup, we start with the initial state as a tensor product of the ancilla qubit  {in the eigenstate $\ket{1}$}  and a spin coherent state, in the tensor product of the Hilbert space  {of the chaos parameter} $\kappa$  and the spin-$j$ Hilbert space. The composite structure of the initial density matrix is given by
\begin{equation}\label{eq: initial density matrix}
    \rho_{\kappa+s}(0) =  \rho_{\kappa}(0) \otimes  \rho_s(0),
\end{equation}
where $\rho_{\kappa}(0) = \ket{1}\bra{1}$ and $\rho_s(0)$ is the density matrix of the spin coherent state as described in Eq.~\eqref{eq:spin coherent state}. 
This state is taken as the input state, as shown in Fig.~\ref{fig:mz_setup}.
\vspace{5mm}

\subsubsection{First beam splitter ($BS_1$)}

The first beam splitter in Fig.~\ref{fig:mz_setup} is denoted by the operator $\hat{M}_{BS_1}$, which is defined by the mixing angles $\sigma_1$ and $\varphi_1$. The angle $\sigma_1$ controls the ratio of transmission ($t_1$) to reflection ($r_1$) of the beam splitter ($BS_1$). The ratio $t_1:r_1$ is given by $(\sin^2(\sigma_1/2) : \cos^2(\sigma_1/2))$ and is tuned by varying $\sigma_1$ in the range $0$ to $\pi$. The matrix representation of $\hat{M}_{BS_1}$ is
\begin{equation}\label{eq:matrix of BS1}
  \hat{M}_{BS_1} =
  \begin{pmatrix}
      \cos(\sigma_1/2)  & e^{-i\varphi_1}\sin(\sigma_1/2) \\
      e^{i\varphi_1}\sin(\sigma_1/2) & -\cos(\sigma_1/2)
  \end{pmatrix}
\end{equation}
and acts on the ancilla qubit with the following transformations:
\begin{align}\label{eq:action of BS1}
     \hat{M}_{BS_1}\ket{1} &= \ket{\psi(\sigma_1,\varphi_1)} \nonumber \\
    &= \cos(\sigma_1/2)\,\ket{1}
    + e^{i\varphi_1}\sin(\sigma_1/2)\,\ket{2}, \\
     \hat{M}_{BS_1}\ket{2} & = |\psi^{\perp}(\sigma_1,\varphi_1)\rangle  \nonumber\\
   & = e^{-i\varphi_1}\sin(\sigma_1/2)\,\ket{1}
    - \cos(\sigma_1/2)\,\ket{2}.
\end{align}
Upon passing through the first beam splitter, the initial state $\rho_{\kappa+s}(0)$ transforms into
\begin{align}
   & \rho'_{\kappa+s}(0) \nonumber \\
    & = \hat{M}_{BS_1}\,\rho_{\kappa+s}(0)\,\hat{M}_{BS_1}^\dagger \nonumber \\
    &= \Big(
        \cos^2(\tfrac{\sigma_1}{2})\,\ket{1}\bra{1}
        + e^{-i\varphi_1}\cos(\tfrac{\sigma_1}{2})\sin(\tfrac{\sigma_1}{2})\,\ket{1}\bra{2} \nonumber \\
    &\quad + e^{i\varphi_1}\cos(\tfrac{\sigma_1}{2})\sin(\tfrac{\sigma_1}{2})\,\ket{2}\bra{1}
        + \sin^2(\tfrac{\sigma_1}{2})\,\ket{2}\bra{2}
      \Big) \otimes \rho_s .
\end{align}
Here the composite state is written in the ancilla basis, that is, the $\kappa$-eigenbasis $\{\ket{1},\ket{2}\}$.

\subsubsection{Time evolution and mirrors} 

 After passing through the first beam splitter $BS_1$, the ancilla is prepared in a superposition of $\ket{1}$ and $\ket{2}$ with amplitudes set by the mixing angle $\sigma_1$. Conditioned on the ancilla state, the spin system undergoes a controlled kicked top evolution (Eq.~\eqref{eq:controlled unitary}), as shown in Fig.~\ref{fig:mz_setup}: along path 1 it evolves under $n$ copies of $U_{\kappa_1}$ and along path 2 under $U_{\kappa_2}$, with chaos parameters $\kappa_1$ and $\kappa_2$, respectively. After $n$ applications of the respective unitary, the joint state takes the block form in the ancilla basis $\{\ket{1},\ket{2}\}$

\begin{widetext}
\begin{equation}
\rho'_{\kappa+s}(n) \;= \;
\begin{pmatrix}
\cos^2(\tfrac{\sigma_1}{2})\,\rho_s^{11}(n) &
\cos(\tfrac{\sigma_1}{2})\sin(\tfrac{\sigma_1}{2})\,e^{-i\varphi_1}\,\rho_s^{12}(n) \\[6pt]
\cos(\tfrac{\sigma_1}{2})\sin(\tfrac{\sigma_1}{2})\,e^{i\varphi_1}\,\rho_s^{21}(n) &
\sin^2(\tfrac{\sigma_1}{2})\,\rho_s^{22}(n)
\end{pmatrix},
\end{equation}
\end{widetext}
where $\rho^{ij}_s(n) = U_{\kappa_i}^n\,\rho_s(0)\,U_{\kappa_j}^{\dagger\,n}$ (with $i,j\in \{1,2\}$) denotes operators acting on the spin Hilbert space.  The off-diagonal terms $\rho_s^{12}$ and $\rho_s^{21}$  result from the interference between the two branches of the superposition. 

\subsubsection{Second beam splitter ($BS_2$) and detection}

 After the controlled evolution, the composite ancilla-spin system enters the second beam splitter, $BS_2$, which acts as a unitary rotation on the ancilla, parameterized by the mixing angle $\sigma_2$ and the phase $\varphi_2$. Its action is described by
\begin{equation}\label{eq:matrix of BS2}
  \hat{M}_{BS_2} =
  \begin{pmatrix}
      \cos(\sigma_2/2)  & e^{-i\varphi_2}\sin(\sigma_2/2) \\
      e^{i\varphi_2}\sin(\sigma_2/2) & -\cos(\sigma_2/2)
  \end{pmatrix}.
\end{equation}

Physically, a click in one of the output ports of the interferometer corresponds to measuring the ancilla in the computational basis   ($\{\ket{1},\ket{2}\}$)  after passing through $BS_2$. Equivalently, this can be described as a measurement on the pre-$BS_2$ state in a rotated basis $\{\ket{D_1},\ket{D_2}\}$, defined by the action of the beam splitter on the computational basis states,
\begin{align}
\ket{D_1(\sigma_2,\varphi_2)} &= \hat{M}_{BS_2}\ket{1} \nonumber \\
&= \cos(\sigma_2/2)\,\ket{1}
+ e^{i\varphi_2}\sin(\sigma_2/2)\,\ket{2}, \nonumber \\
\ket{D_2(\sigma_2,\varphi_2)} &= \hat{M}_{BS_2}\ket{2} \nonumber \\
&= e^{i\varphi_2}\sin(\sigma_2/2)\,\ket{1}
- \cos(\sigma_2/2)\,\ket{2}.
\end{align}
A measurement click in a given detector corresponds to projecting the composite ancilla–spin state onto the corresponding rotated ancilla state, $\ket{D_1}$ or $\ket{D_2}$, respectively. Since the measurement acts only on the ancilla subsystem, the full measurement operators are constructed by taking the outer product $|D_i\rangle\langle D_i|$ on the ancilla space and the identity operator $I_s$ on the spin space. The resulting operators define the projective measurement associated with each detector  and are given by 
\begin{align}\label{eq:projection operator}
P_1 &= \ket{D_1}\bra{D_1}\otimes I_s,\\
P_2 &= \ket{D_2}\bra{D_2}\otimes I_s.
\end{align} 

Rather than explicitly applying $\hat{M}_{BS_2}$ to the joint state, we adopt this equivalent description and incorporate the action of $BS_2$ into the measurement projectors. Since we are interested in the interference between the two evolutions, we post-select, without loss of generality, on the outcome associated with port~$D_1$. The resulting post-selected (unnormalized) spin state is obtained by projecting the pre-$BS_2$ joint state onto $\ket{D_1}$ and tracing out the ancilla,
\begin{align}\label{eq: unormalized post selected spin density}
& \tilde\rho^{\mathrm{ps}}_s(n;\sigma_1,\varphi_1;\sigma_2,\varphi_2) \nonumber \\
&= \operatorname{Tr}_{\kappa}\!\left[P_1\,\rho'_{\kappa+s}(n)\right] \nonumber \\
& = \bra{D_1}\rho'_{\kappa+s}(n)\ket{D_1} \nonumber \\
&= \cos^2(\tfrac{\sigma_2}{2})\cos^2(\tfrac{\sigma_1}{2})\rho_s^{11}(n) + \sin^2(\tfrac{\sigma_2}{2})\sin^2(\tfrac{\sigma_1}{2})\,\rho_s^{22}(n) \nonumber \\
   & +\frac{1}{4}\sin(\sigma_2)\sin(\sigma_1)\bigg(e^{i(\varphi_2-\varphi_1)}\rho_s^{12}(n) + e^{-i(\varphi_2-\varphi_1)}\rho_s^{21}(n)\bigg) \nonumber \\
\end{align}
and the normalized state is given by
\begin{equation}\label{eq: normalized post selected spin density}
{\rho}^{\mathrm{ps}}_s(n;\sigma_1,\varphi_1;\sigma_2,\varphi_2)
=
\frac{\tilde\rho^{\mathrm{ps}}_s(n;\sigma_1,\varphi_1;\sigma_2,\varphi_2)}
{\operatorname{Tr}\!\left[\tilde\rho^{\mathrm{ps}}_s(n;\sigma_1,\varphi_1;\sigma_2,\varphi_2)\right]}.
\end{equation}
The corresponding detection probability is
\begin{equation}
p_1 =
\operatorname{Tr}_s\!\left[\tilde\rho^{\mathrm{ps}}_s(n;\sigma_1,\varphi_1;\sigma_2,\varphi_2)\right].
\end{equation}

 Since the two paths are distinguished by different chaos parameters $\kappa_{1,2}$, they can realize genuinely different dynamical evolutions, in contrast to a conventional  Mach-Zehnder interferometer, in which the two paths differ only by a relative phase. For mathematical simplicity, we therefore set $\varphi_1=\varphi_2=0$ in Eq.~\eqref{eq: normalized post selected spin density}, as our primary interest is to study the interference between two classically distinct dynamical evolutions. 

\subsubsection{Classical mixture}

Alternatively, instead of post-selecting on a particular measurement outcome, we may perform a non-selective measurement, i.e., trace  over the ancilla after the second beam splitter. This yields the unconditional spin state, which corresponds to a classical mixture of the two kicked top evolutions with  chaos parameters $\kappa_1$ and $\kappa_2$:
\begin{align}\label{eq: classical spin density}
\rho_s^{\mathrm{cl}}(n;\sigma_1,\varphi_1) 
&= \operatorname{Tr}_{\kappa}\!\left[\rho''_{\kappa+s}(n;\sigma_1,\varphi_1 {=0};\sigma_2,\varphi_2 {=0})\right] \nonumber \\
&= \operatorname{Tr}_{\kappa}\!\bigl[ \ket{D_1}\bra{D_1} \otimes \cos^2{(\sigma_1/2)}\,  \rho^{11}_s  \nonumber \\
&\qquad + \ket{D_2}\bra{D_2} \otimes \sin^2({\sigma_1/2}) \,\rho^{22}_s \bigr] \nonumber \\
&=  \cos^2{(\sigma_1/2)}\,  \rho^{11}_s + \sin^2({\sigma_1/2}) \,\rho^{22}_s,
\end{align}
where $\rho''_{\kappa+s}$ is the 
density matrix of the system after emerging from the second beam splitter. This indicates 
that all interference terms cancel when both detectors are included. Physically, this corresponds to evolving the system under two kicked top dynamics with parameters $\kappa_1$ and $\kappa_2$, and then taking a classical mixture of the two resulting states, with weights determined by the mixing angle $\sigma_1$ of the first beam splitter. In this non-selective case, the second beam splitter does not affect the final density matrix, since tracing out the ancilla removes all coherence between the two paths.


\section{Observations}

\begin{figure*}[t]
\centering
\begin{tabular}{ccc} 
\includegraphics[trim={90mm 5mm 100mm 13mm},clip,width=0.33\textwidth]{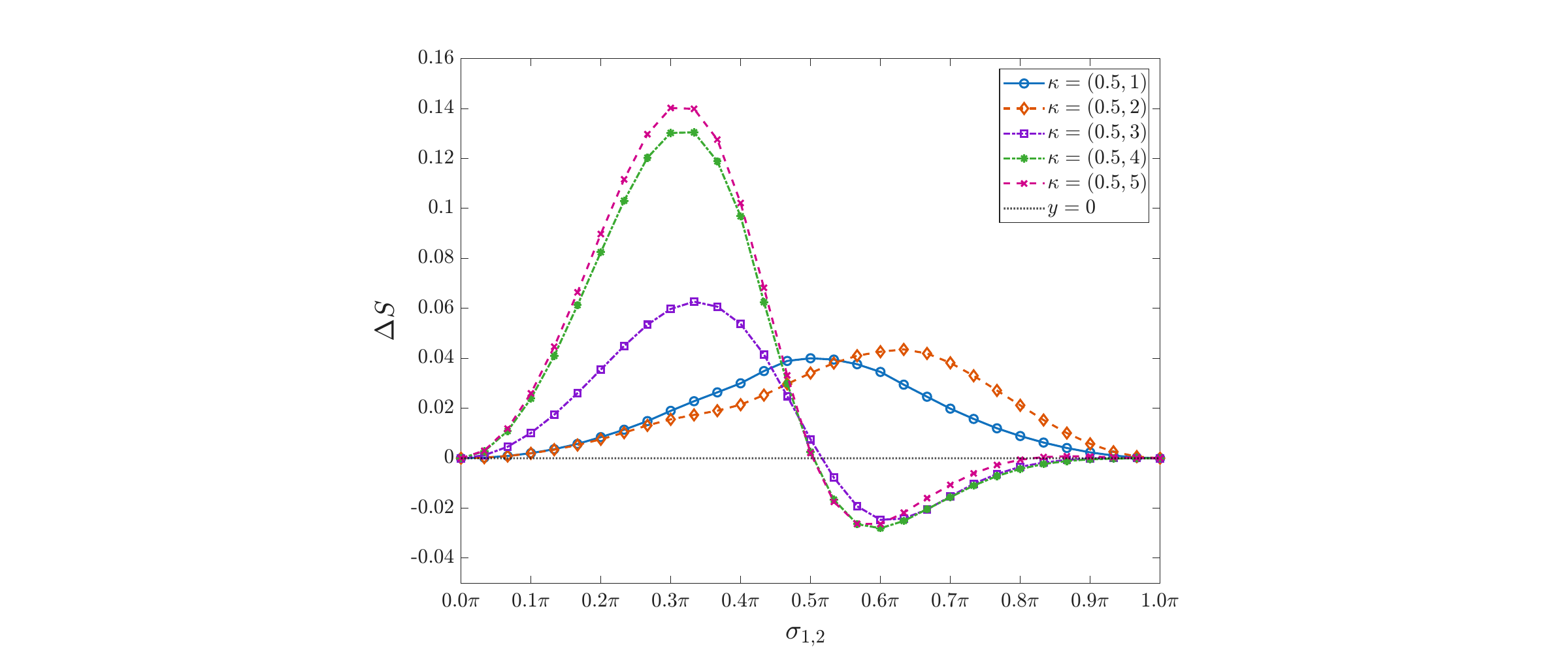}\label{fig:entropy_diff_vs_bloch_angle_1} &
\includegraphics[trim={90mm 5mm 100mm 13mm},clip,width=0.33\textwidth]{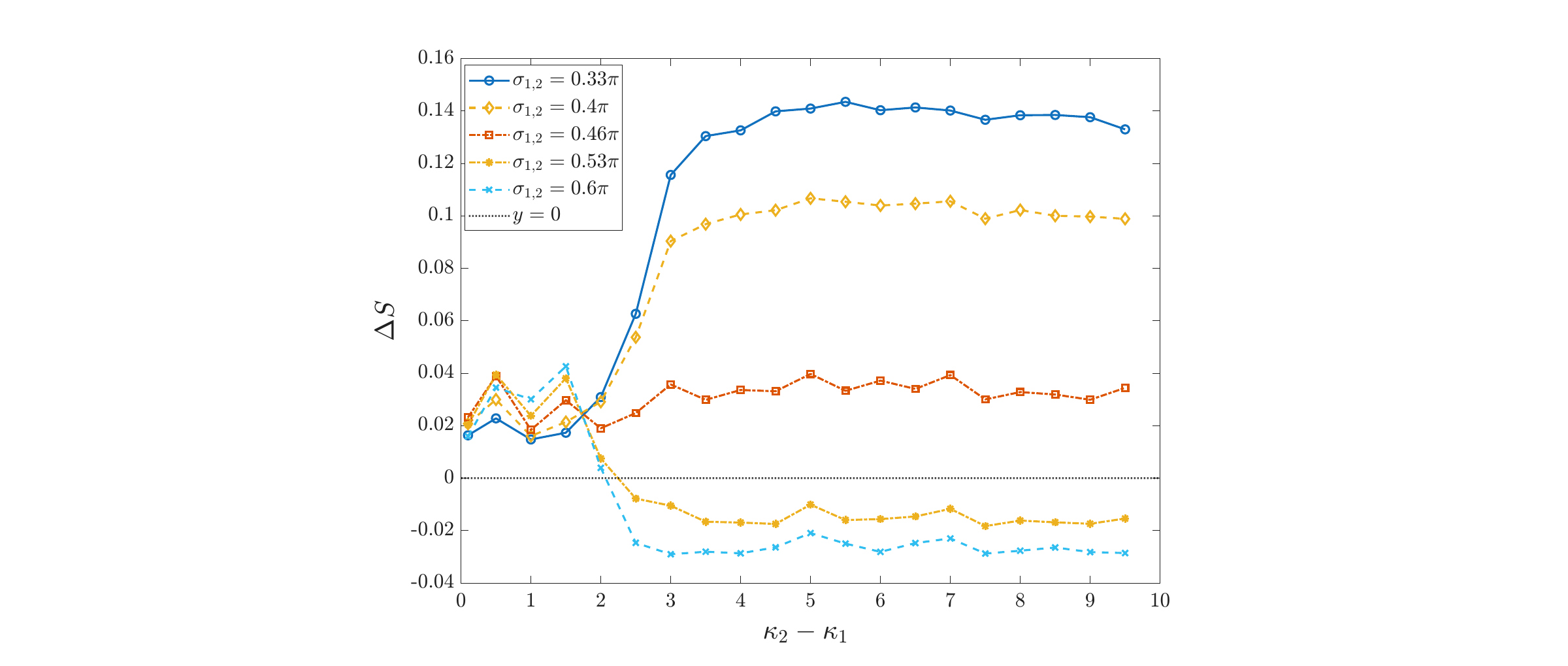}\label{fig:entropy_diff_vs_kappa_diff_1} &
\includegraphics[trim={87mm 4mm 94mm 13mm},clip,width=0.335\textwidth]{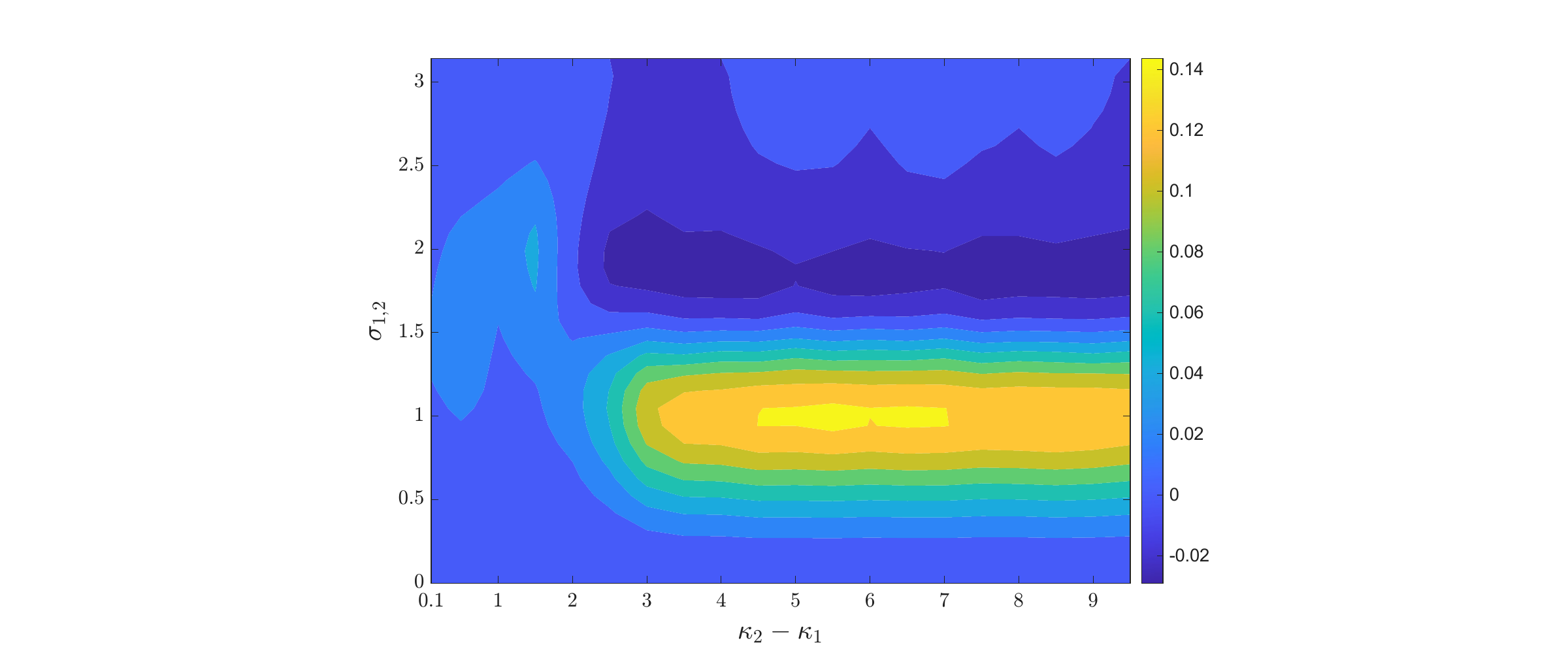}\label{fig:entropy_diff_heatmap_1} \\
(a) & (b) & (c)
\end{tabular}
\caption{Time-averaged subsystem entropy difference $\Delta S$, averaged over 200 periods of the kicked top unitary. The spin coherent state is initialized at $(\theta,\phi) = (2.25, 1.1)$, while the ancilla qubit is initialized in the state $\ket1$. (a) $\Delta S$ as a function of the beam splitter angles, with $\sigma_1=\sigma_2$, with 31 points in the interval $(0,\pi)$. Different curves correspond to different pairs of  chaos parameters  $\kappa$, with $\kappa_1 = 0.5$ (regular dynamics) fixed and $\kappa_2 $ varied from $1$ to $5$. (b) $\Delta S$ as a function of the  chaos parameter difference $\kappa_2-\kappa_1$, sampled with 20 points in the interval $(0.1,9.5)$. Different curves correspond to different beam splitter angles $\sigma_1 = \sigma_2$. (c) Contour plot showing the dependence of $\Delta S$ on the chaos parameter difference $\kappa_2-\kappa_1$  and the beam splitter angle $\sigma_{1,2}$, sampled with 20 and 31 points, respectively. The black dotted line indicates $\Delta S=0$ in panels (a) and (b).}
\label{fig:entropy_diff_vs_kappa_sigma_1}
\end{figure*}

\begin{figure*}[t]
\centering
\begin{tabular}{ccc} 
\includegraphics[trim={90mm 5mm 100mm 13mm},clip,width=0.33\textwidth]{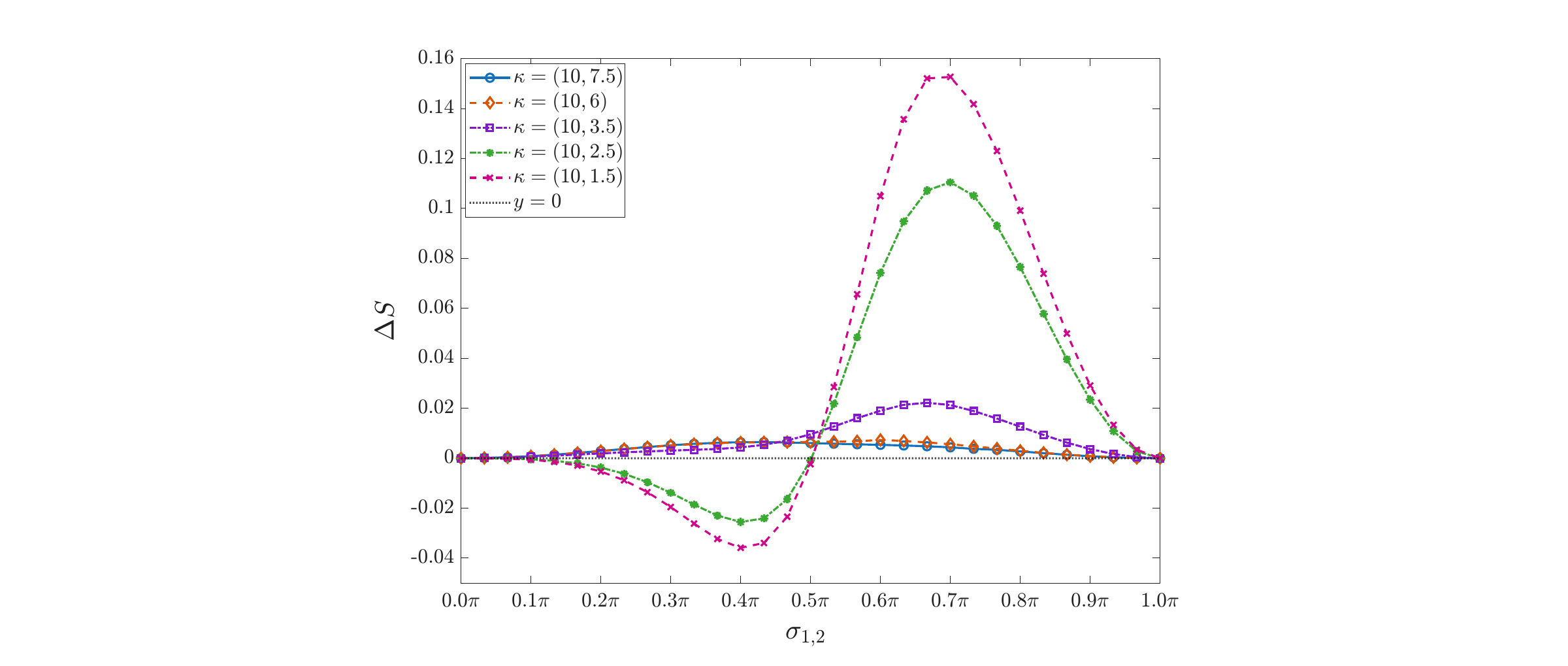}\label{fig:entropy_diff_vs_bloch_angle3} &
\includegraphics[trim={90mm 5mm 100mm 13mm},clip,width=0.33\textwidth]{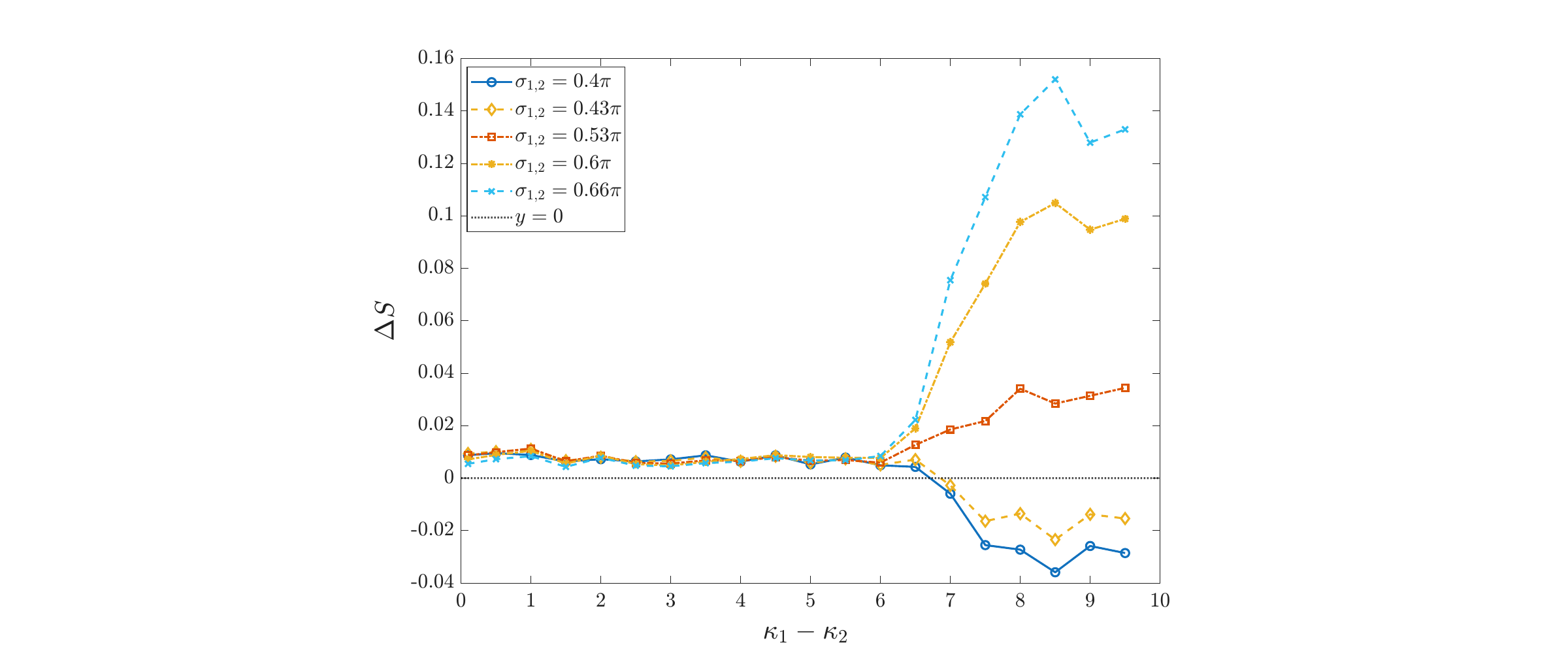}\label{fig:entropy_diff_vs_kappa_diff3} &
\includegraphics[trim={87mm 4mm 94mm 13mm},clip,width=0.335\textwidth]{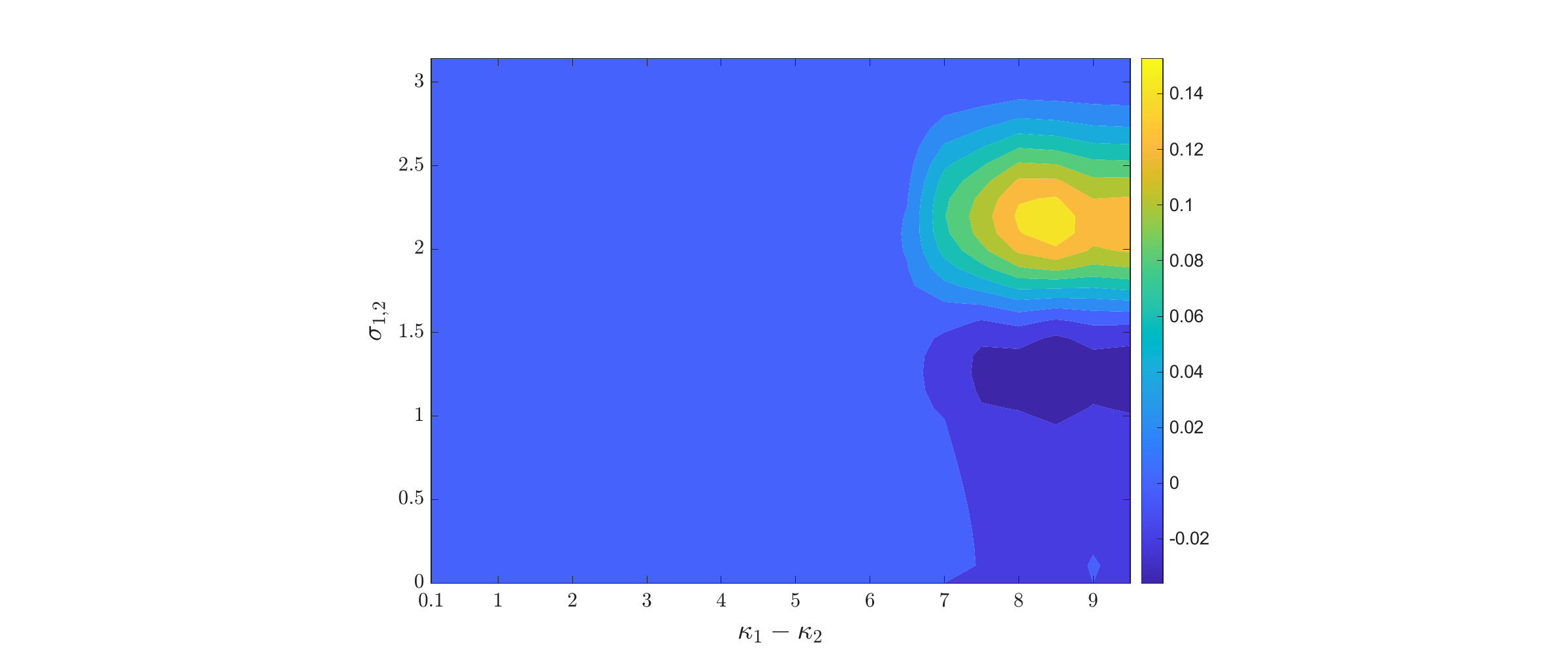}\label{fig:entropy_diff_heatmap3} \\
(a) & (b) & (c)
\end{tabular}
\caption{Time-averaged subsystem entropy difference $\Delta S$, averaged over 200 periods of the kicked top unitary. The spin coherent state is initialized at $(\theta,\phi) = (2.25, 1.1)$, and the ancilla qubit is initialized in the state  $\ket1$. (a) $\Delta S$ as a function of the beam splitter angles, with $\sigma_1=\sigma_2$, sampled with 31 points in the interval $(0,\pi)$. Different curves correspond to different pairs of  chaos parameters  $\kappa$, with $\kappa_1 = 10$ (chaotic dynamics) fixed and $\kappa_2 $ varied from $7.5$ to $1.5$. (b) $\Delta S$ as a function of the  chaos parameters difference $\kappa_1-\kappa_2$, sampled with 20 points in the interval $(0.1,9.5)$. Different curves correspond to different beam splitter angles $\sigma_1 = \sigma_2$. (c) Contour plot showing the dependence of $\Delta S$ on the  chaos parameters difference $\kappa_1-\kappa_2$ and the beam splitter angle $\sigma_{1,2}$, sampled with 20 and 31 points, respectively. The black dotted line indicates $\Delta S=0$ in panels (a) and (b).}
\label{fig:entropy_diff_vs_kappa_sigma_2}
\end{figure*}

We shall employ the von Neumann entropy of a subsystem of the spin system as a diagnostic to study the difference between two types of quantum evolutions: one obtained through post-selection and the other corresponding to a classical mixture. In previous studies of the quantum kicked top, the von Neumann entropy has been widely used as a dynamical signature of the transition from regular to chaotic behaviour as the parameter $\kappa$ is varied \cite{robert_kicked_top_1996,untangling_ghose_meenu_2019, kicked_top_silvia_2020}. The quantum kicked top has a natural representation in terms of spin-$1/2$ particles (qubits): A spin-$j$ kicked top can be mapped to a system of $N = 2j$ qubits confined to the symmetric subspace of the full Hilbert space $\mathbb{C}^{2^N}$. Throughout this paper, we use the single-qubit von Neumann entropy, defined as
\begin{equation}
S(\rho) = -\operatorname{Tr}\!\left(\rho \log \rho \right),
\end{equation}
where $\rho$ is the reduced density matrix obtained from the spin system by tracing out $N-1$ qubits. The entropy ranges from $0$ (the single-qubit state is pure) to $\log 2$ (the single qubit state is maximally mixed). The quantity of interest in our analysis is the difference between the single-qubit entropy in the classical mixture case and the post-selected case, defined as
\begin{equation}\label{eq:single qubit entropy difference}
\Delta S(n) = S \big( \operatorname{Tr}_{N-1}(\rho_s^{\mathrm{cl}}) \big) - S \big( \operatorname{Tr}_{N-1}({\rho}_s^{\mathrm{ps}}) \big),
\end{equation}
where ${\rho}_s^{\mathrm{ps}}$ and $\rho_s^{\mathrm{cl}}$ are the post-selected and classical spin density matrices defined in Eqs.~\eqref{eq: normalized post selected spin density} and \eqref{eq: classical spin density}, respectively.

\subsection{ Single initial state with $\sigma_1=\sigma_2$}

In this section, we analyze the entropy difference defined in Eq.~\eqref{eq:single qubit entropy difference}, averaged over a fixed number of periods, for the case in which both beam splitters have identical  mixing angles,  i.e., $\sigma_1 = \sigma_2 = \sigma$ and $\varphi_1 = \varphi_2 = 0$. Under these conditions, the normalized post-selected state takes the form
\begin{align}\label{eq: normalized post selected spin density for same sigma}
&{\rho}^{\mathrm{ps}}_s(n;\sigma) \nonumber \\
= &\frac{\tilde\rho^{\mathrm{ps}}_s(n;\sigma)}
     {\operatorname{Tr}\!\left[\tilde\rho^{\mathrm{ps}}_s(n;\sigma)\right]}\nonumber \\[4pt]
= &\frac{1}{\operatorname{Tr}\!\left[\tilde\rho^{\mathrm{ps}}_s(n;\sigma)\right]}\big(\cos^4\!\Big(\tfrac{\sigma}{2}\Big)\,\rho_s^{11}(n) +\sin^4\!\Big(\tfrac{\sigma}{2}\Big)\,\rho_s^{22}(n) \nonumber   \\[-2pt]
&\qquad + \tfrac{1}{4}\sin^2(\sigma)\big(\rho_s^{12}(n) + \rho_s^{21}(n)\big)\big).
\end{align}
 We begin with a spin coherent state pointing in the direction  $(\theta = 2.25,\phi = 1.1)$, defined in  Eq.~\eqref{eq:spin coherent state}, with spin size of $j=25$. This state is chosen because, in the classical limit of the quantum kicked top, it undergoes a transition from regular to chaotic dynamics as the chaos parameter $\kappa$ increases, with the threshold occurring near $\kappa_{\mathrm{th}} \approx 2.5$. By selecting pairs of $\kappa$ values both below and above this threshold, we can assess how classical chaos influences the superposed evolution. 

The ancilla qubit is initialized in the state $\ket{1}$. The composite state is then evolved using the Mach–Zehnder interferometric setup for various pairs of $\kappa$ values, as shown in Fig.~\ref{fig:entropy_diff_vs_kappa_sigma_1}. For each pair, we compute the classical mixture and post-selected spin density matrices using Eqs.~\eqref{eq: classical spin density} and \eqref{eq: normalized post selected spin density for same sigma}, and evaluate the corresponding single-qubit entropy difference $\Delta S(n)$ defined in Eq.~\eqref{eq:single qubit entropy difference}.

For each pair of chaos parameters $(\kappa_1,\kappa_2)$, we calculate $\Delta S(n)$ over $n=0,\dots,200$ kicked top periods and use its time average as the figure of merit.  For the rest of the paper we will denote the time-averaged subsystem entropy difference as $\Delta S$.  This is repeated for different beam splitter angles  $\sigma$, which determine the relative weights of the two unitary evolutions and the measurement settings.

The results, shown in Fig.~\ref{fig:entropy_diff_vs_kappa_sigma_1}(a), indicate that the subsystem entropy difference $\Delta S > 0$ when both  chaos parameters $\kappa_1$ and $\kappa_2$ correspond to regular dynamics. As  the value of  $\kappa_2$ increases with $\kappa_1 = 0.5$ fixed, $\Delta S$ becomes negative for $\sigma > \pi/2$ once $\kappa_2$ exceeds the threshold $\kappa_{\mathrm{th}}$; in this regime, the superposition is  weighted more strongly toward the more chaotic evolution. For $\sigma \leq \pi/2$, $\Delta S$ remains positive and grows with $\kappa_2$, reflecting increased contrast between the superposed and classically mixed evolutions. As expected, $\Delta S = 0$ at $\sigma = 0$ or $\pi$, where the dynamics reduces to a single kicked top evolution.

In Fig.~\ref{fig:entropy_diff_vs_kappa_sigma_1}(b), we show the dependence of the time-averaged single-qubit von Neumann entropy difference $\Delta S$ on the difference  between the two $\kappa$ values. Here, $\kappa_1 = 0.5$ is fixed and 
$\kappa_2$ is varied. The plotted entropy difference, averaged over 200 kicks, is shown as a function of the  difference $\kappa_2-\kappa_1$. For differences less than or equal to $2$, the entropy difference remains small and positive for all values of beam splitter angles $\sigma$. Once the difference in chaos parameters exceeds $2$ (corresponding to $\kappa_2 \gtrsim 2.5$), the curve dips below zero for $\sigma \approx   0.53\pi$, which is greater than $\pi/2$. For all $\sigma > \pi/2$ (not shown here), $\Delta S$ remains negative for $\kappa_2 \gtrsim 2.5$. For $\sigma \leq \pi/2$, the entropy difference increases with increasing ($\kappa_2-\kappa_1$), indicating that the subsystem entropy is always larger for the classical mixture than for the superposition in this regime.

In Fig.~\ref{fig:entropy_diff_vs_kappa_sigma_1}(c), we present a heat map of the subsystem entropy difference $\Delta S$ as a function of $\sigma$ (ranging from $0$ to $\pi$) and the difference  $\kappa_2-\kappa_1$ (from $0.1$ to $9.5$). The heat map shows that negative values of $\Delta S$ occur only in a narrow region characterized by $\sigma > \pi/2$ and $\kappa$ differences greater than $2$. These observations collectively indicate that, in order for the subsystem entropy of the superposed evolution to exceed that of  the classical mixture, one of the evolutions must be chaotic ($\kappa_i \gtrsim 2.5$), and for that chaotic evolution to have greater weight in the superposition. We further verified this behaviour using a different initial spin coherent state, $(\theta,\phi) = (2.25,-1.6)$, which also exhibits a regular-to-chaotic transition with increasing  chaos parameter  $\kappa$. The results show the same qualitative features; see  {Fig.~\ref{fig:entropy_diff_vs_kappa_sigma_3} in} the Supplementary Material for details.

In Fig.~\ref{fig:entropy_diff_vs_kappa_sigma_2}, we perform a similar analysis using the same initial state, $(\theta,\phi) = (2.25,1.1)$, but now starting with a pair of highly chaotic $\kappa$ values. In this case, we fix $\kappa_1 = 10$, for which the kicked top is strongly chaotic in the classical limit, and gradually decrease $\kappa_2$ from the chaotic regime toward the regular regime. In Fig.~\ref{fig:entropy_diff_vs_kappa_sigma_2}(a), the entropy difference $\Delta S$ remains positive as long as both $\kappa$ values lie in the chaotic regime. However, once $\kappa_2$ moves into the regular  regime, we observe a drop in $\Delta S$ for $\sigma < \pi/2$, where the branch associated with $\kappa_2$ carries greater weight in the superposition. This behaviour is consistent with Fig.~\ref{fig:entropy_diff_vs_kappa_sigma_2}(b), where the subsystem entropy difference $\Delta S$ is plotted as a function of the difference $\kappa_1-\kappa_2$  for various values of beam splitter angles $\sigma$. Here, again, we find that when the $\kappa$ difference becomes large enough for the second parameter to enter the regular regime, the entropy difference becomes negative for smaller values of $\sigma$. 

Similar to the previous case, the heat map shown in Fig.~\ref{fig:entropy_diff_vs_kappa_sigma_2}(c) reveals a small region in which $\Delta S$ is negative. This occurs only when $\sigma < \pi/2$ and the two $\kappa$ values differ sufficiently for one of them to lie in the regular regime. These results indicate that when two chaotic or two regular evolutions are superposed, the resulting time-averaged subsystem entropy is less than that of the corresponding classical mixture, regardless of   the relative amplitudes in the superposition. In contrast, a higher time-averaged subsystem entropy in the superposition compared to the classical mixture appears only when the two evolutions are dynamically distinct (one chaotic and one regular), with the chaotic evolution having greater weight in the superposition.

\subsection{Different initial states with $\sigma_1 = \sigma_2$} \label{sec: Different initial states, same sigma}

\begin{figure*}[t]
\centering

\begin{subfigure}[t]{0.32\textwidth}
\centering
\includegraphics[trim={100mm 8mm 100mm 13mm},clip,width=\textwidth]{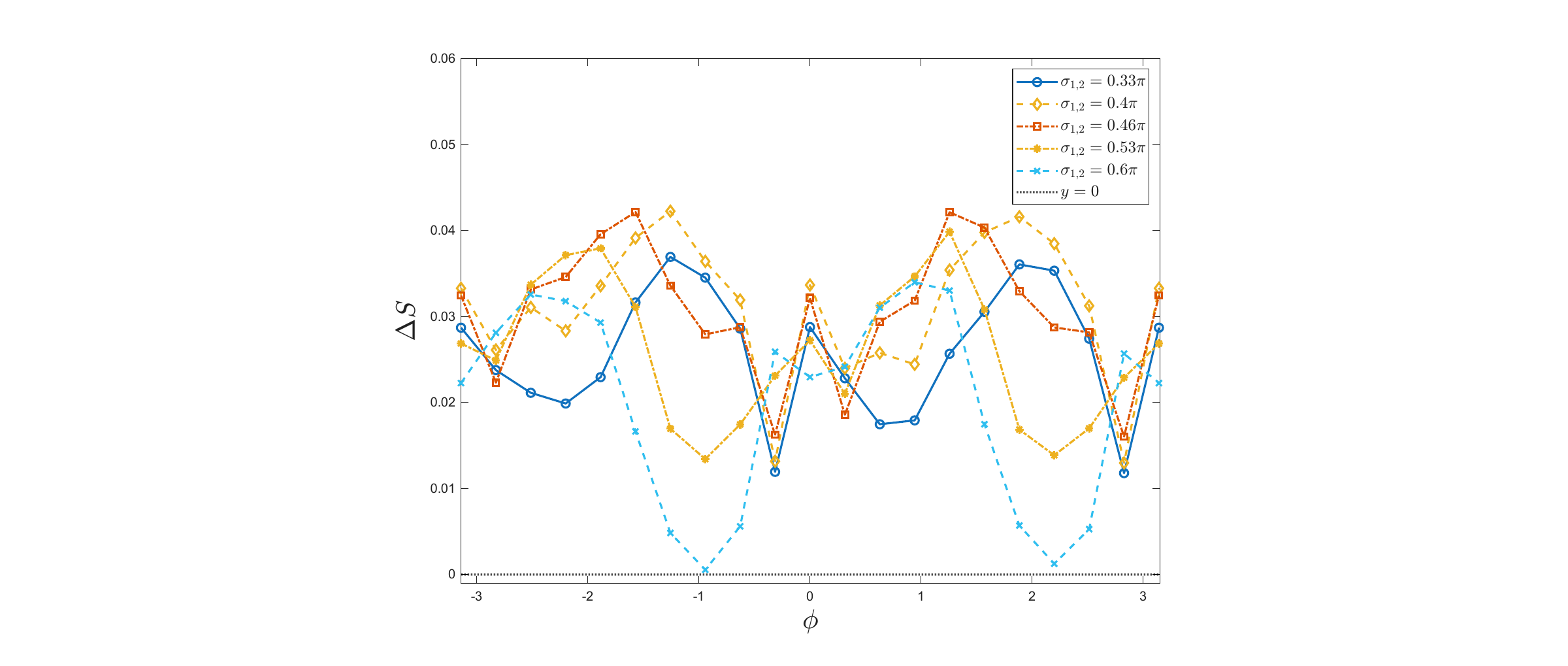}
\caption{$(\kappa_1,\kappa_2)=(0.5,1.0)$}
\label{fig:entropy_vs_phi_1}
\end{subfigure}
\hfill
\begin{subfigure}[t]{0.32\textwidth}
\centering
\includegraphics[trim={100mm 8mm 100mm 13mm},clip,width=\textwidth]{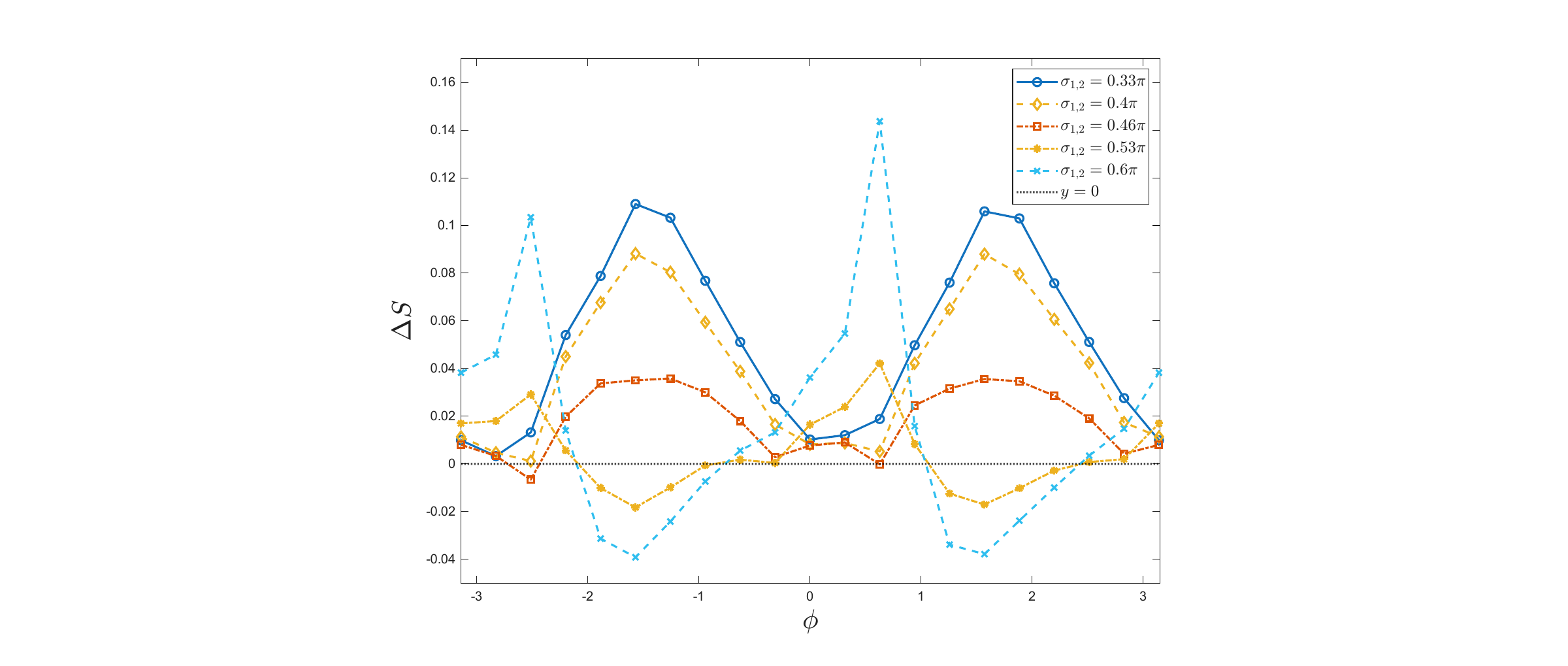}
\caption{$(\kappa_1,\kappa_2)=(0.5,3.0)$}
\label{fig:entropy_vs_phi_2}
\end{subfigure}
\hfill
\begin{subfigure}[t]{0.32\textwidth}
\centering
\includegraphics[trim={100mm 8mm 100mm 13mm},clip,width=\textwidth]{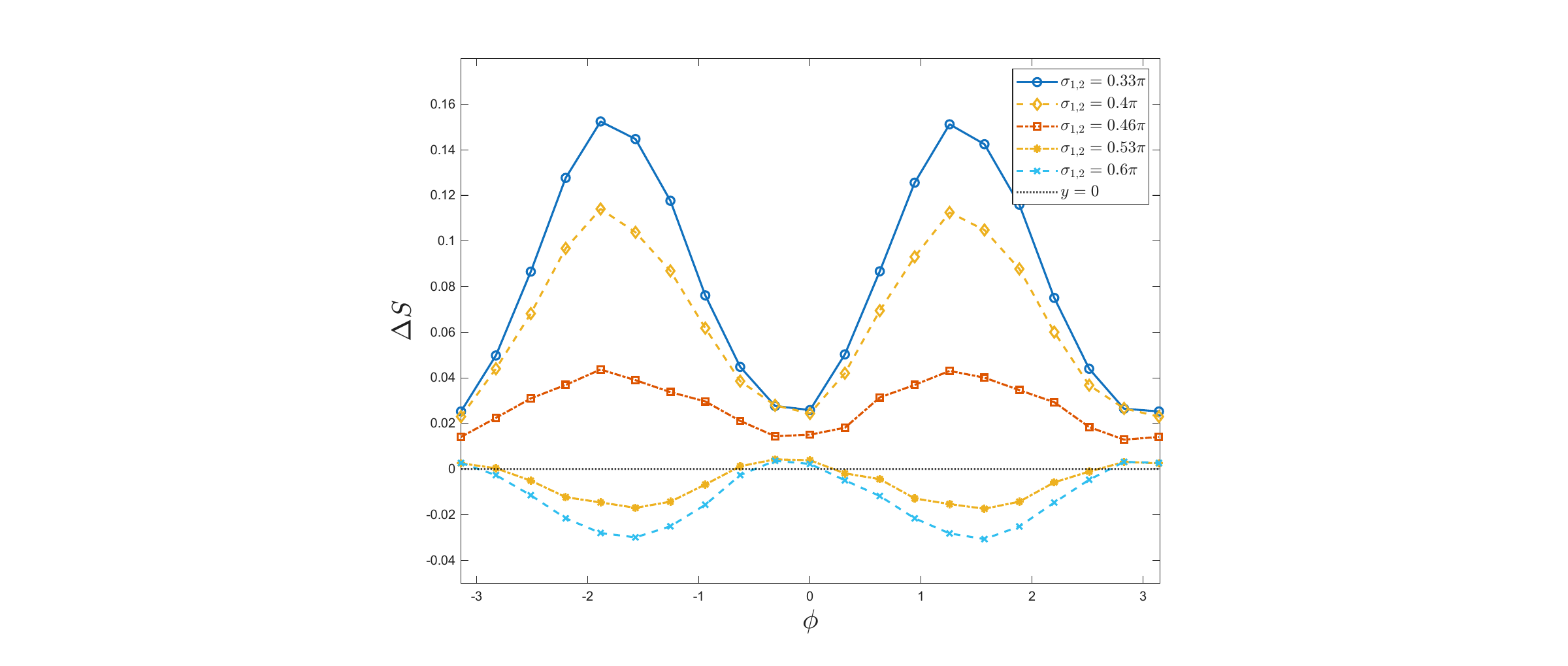}
\caption{$(\kappa_1,\kappa_2)=(0.5,6.0)$}
\label{fig:entropy_vs_phi_3}
\end{subfigure}

\caption{
Time-averaged subsystem entropy difference $\Delta S$, defined in Eq.~\eqref{eq:single qubit entropy difference}, averaged over 200 periods of the kicked top unitary for three pairs of  chaos parameters $(\kappa_1,\kappa_2)$.
Each curve corresponds to a different beam splitter mixing angle with $\sigma_1=\sigma_2$.
For each curve, we sample 21 spin coherent initial states at fixed polar angle $\theta=2.25$ and uniformly spaced azimuthal angles $\phi\in[-\pi,\pi]$.
These initial conditions correspond to the states lying on the black dotted line in the classical stroboscopic maps of Fig.~\ref{fig:classical_map}.
The black dotted horizontal line in these figures indicates $\Delta S=0$.
The spin size is $j=25$.
}
\label{fig:entropy_diff_vs_phi}
\end{figure*}

In this subsection we study different initial states for the spin system  with spin value $j=25$ and compare their behaviour directly with the corresponding classical phase space structure. Figures~\ref{fig:classical_0.5} -- \ref{fig:classical_6} show the global classical phase space of the classical kicked top for  chaos parameters $\kappa = 0.5, 1, 3,$ and $6$, respectively. For each plot, 289 initial points parametrized by $(\theta,\phi)$ were evolved using the stroboscopic equations of motion in Eq.~\eqref{eq:classical_eom} for 400 kicks.  
For $\kappa = 0.5$ and $1$, the phase space is almost entirely regular. As  the value of $\kappa$ increases to $3$, the regular islands shrink and a large chaotic sea emerges, giving rise to a mixed phase space. For $\kappa = 6$, the phase space becomes almost fully chaotic, with only a few isolated periodic points remaining.

To connect these classical structures with our quantum analysis, we fix $\theta = 2.25$ and vary $\phi$ in the range $[-\pi,\pi]$ (see the dotted black line in Figs.~\ref{fig:classical_0.5}–\ref{fig:classical_6}). This allows us to probe initial states that traverse different dynamical regions -- regular islands, chaotic seas, and intermediate zones -- using a one-dimensional slice of phase space. Similar approaches have been previously employed to investigate the connection between classical chaos and entanglement growth \cite{wang_ghose_entg_2004,Ghose_chaos_ent_dec_2008,chaudhury_2009}.  

We choose 21 evenly spaced values of $\phi$ in the interval $[-\pi,\pi]$ with fixed $\theta = 2.25$. Each of the 21 spin coherent states, tensored with the ancilla qubit, is evolved under the Mach–Zehnder interferometric setup with the quantum kicked top for 200 kicks, for various beam splitter angles $\sigma_1=\sigma_2=\sigma$ and $\varphi_1 = \varphi_2 = 0$.  
As in the previous section, we compute both the classical-mixture and post-selected spin density matrices. We then evaluate the subsystem entropy difference $\Delta S(n)$ defined in Eq.~\eqref{eq:single qubit entropy difference} for all 21 initial states and different values of $\sigma$. Then we take the time-average over 200 time periods to get $\Delta S$. This analysis is repeated for three different pairs of chaos parameters $(\kappa_1,\kappa_2)$: $(0.5,1)$, $(0.5,3)$, and $(0.5,6)$, as shown in Figs.~\ref{fig:entropy_vs_phi_1}, \ref{fig:entropy_vs_phi_2}, and \ref{fig:entropy_vs_phi_3}.

Fig.~\ref{fig:entropy_vs_phi_1} corresponds to a pair of $\kappa$ values that both lie in the regular regime, as seen in Figs.~\ref{fig:classical_0.5} and \ref{fig:classical_1}.  
Here, the subsystem entropy difference $\Delta S$ shows no significant dependence on $\phi$: the distinction between the superposed and classical-mixture evolutions remains small for all beam splitter angles  $\sigma$, with only mild oscillations across the interval $[-\pi,\pi]$.

In the next case, we superpose evolutions with $\kappa_1 = 0.5$ (regular) and $\kappa_2 = 3$ (mixed).  
For $\kappa = 3$, varying $\phi$ moves the initial state across regular islands and chaotic regions, as seen in Fig.~\ref{fig:classical_3}. In Fig.~\ref{fig:entropy_vs_phi_2}, the subsystem entropy difference becomes negative for $\sigma > \pi/2$ precisely in those $\phi$ intervals where the classical dynamics is chaotic. In contrast, for $\sigma < \pi/2$, $\Delta S$ remains positive (with a small exception at $\sigma=0.46\pi$). This indicates that the superposition of one chaotic and one regular evolution yields higher subsystem entropy than the classical mixture when the chaotic component has sufficiently large weight.

This effect becomes even more pronounced when we combine a highly chaotic regime ($\kappa_2 = 6$) with a regular one ($\kappa_1 = 0.5$). For $\kappa = 6$, the dynamics is fully chaotic over the entire $\phi$ range (Fig.~\ref{fig:classical_6}). In Fig.~\ref{fig:entropy_vs_phi_3}, the subsystem entropy difference $\Delta S$ shows a clear separation between curves for different $\sigma$ values: for $\sigma > \pi/2$, $\Delta S$ remains negative for all $\phi$, while for $\sigma \leq \pi/2$ it stays positive. A small region near $\phi = 0$ shows positive values for $\sigma>\pi/2$, likely due to finite-size effects and residual influence of the regular $\kappa_1=0.5$ dynamics. These finite size effects can be removed by considering a higher spin value such as $j=200$, as shown in the Supplementary Material (see Fig.~\ref{fig:entropy_diff_vs_phi_2}(c)).
Overall, the trend is consistent: superposing a chaotic and a regular evolution leads to greater subsystem entropy generation than the classical mixture when the chaotic contribution dominates.

\subsection{ Single initial state with $\sigma_1\neq \sigma_2$}

In this section, we study how different beam splitter mixing angles affect the superposed quantum evolution. We fix the initial spin state with spin value $j=25$ and, for each pair of $\kappa$ values, we vary the beam splitter angles $\sigma_{1,2} \in [0,\pi]$ using 31 evenly spaced values. For a given chaos parameter pair ($\kappa_1,\kappa_2$), the spin system is initialized in the spin coherent state pointing in the direction $(\theta,\phi) = (2.25,1.1)$, while the ancilla qubit is initialized in the state $\ket{1}$. After passing through the first beam splitter,  parametrized by $\sigma_1$, the composite state evolves through the quantum kicked top unitary for $n$ time periods: the evolution in path~1 uses $U_{\kappa_1}$, and in path~2 uses $U_{\kappa_2}$.
We then vary the angle of the second beam splitter $\sigma_2$ and apply the projector defined in Eq.~\eqref{eq:projection operator}. The corresponding post--selected spin state for each $(\sigma_1,\sigma_2)$ pair is given in Eq.~\eqref{eq: normalized post selected spin density}. For the classical mixture of the two evolutions, the second beam splitter has no effect on the final spin density matrix, since it acts only on the ancilla, which is subsequently traced out.

\begin{figure*}
\centering
\begin{tabular}{ccc} 
\begin{subfigure}{0.33\textwidth}
    \centering
    \includegraphics[width=\textwidth]{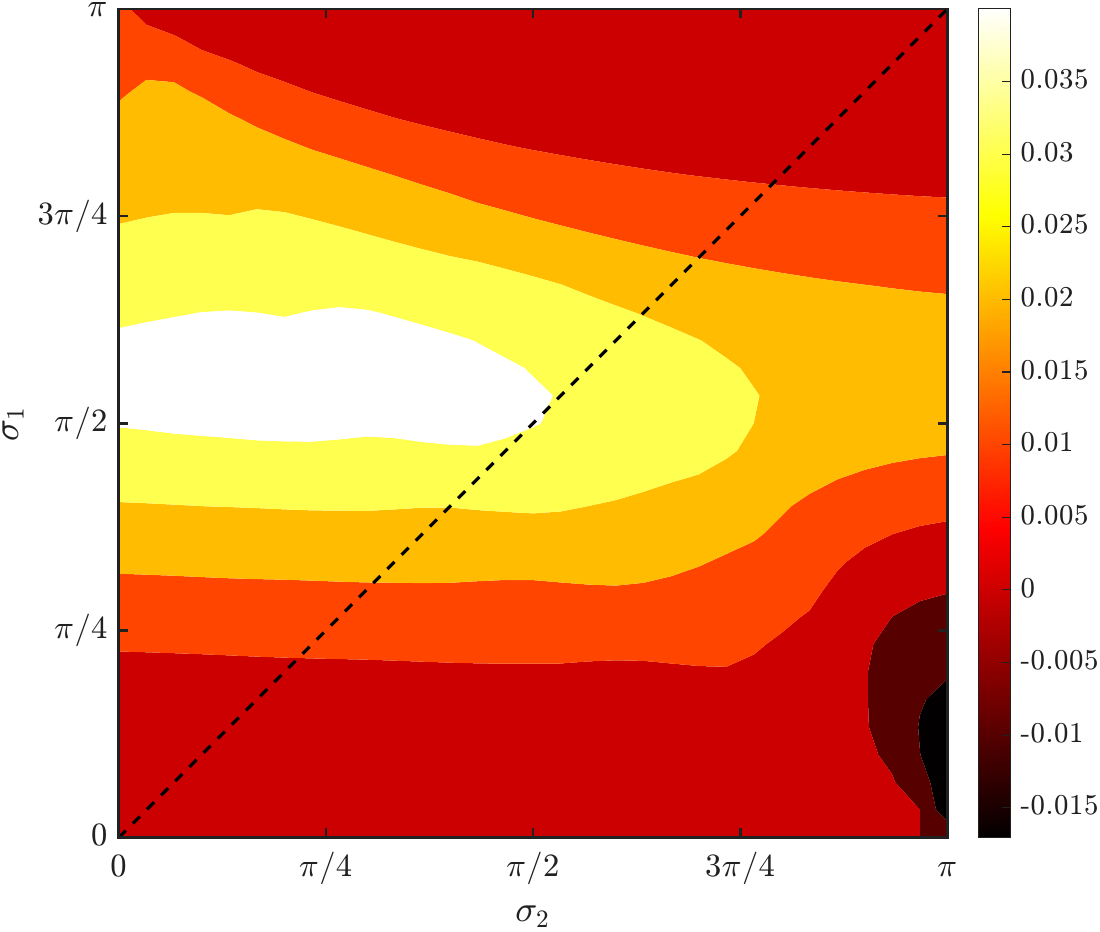}
    \caption{$(\kappa_1,\kappa_2) =(0.5,1)$}
    \label{fig:contour_sigma_1}
\end{subfigure}
&
\begin{subfigure}{0.33\textwidth}
    \centering
    \includegraphics[width=\textwidth]{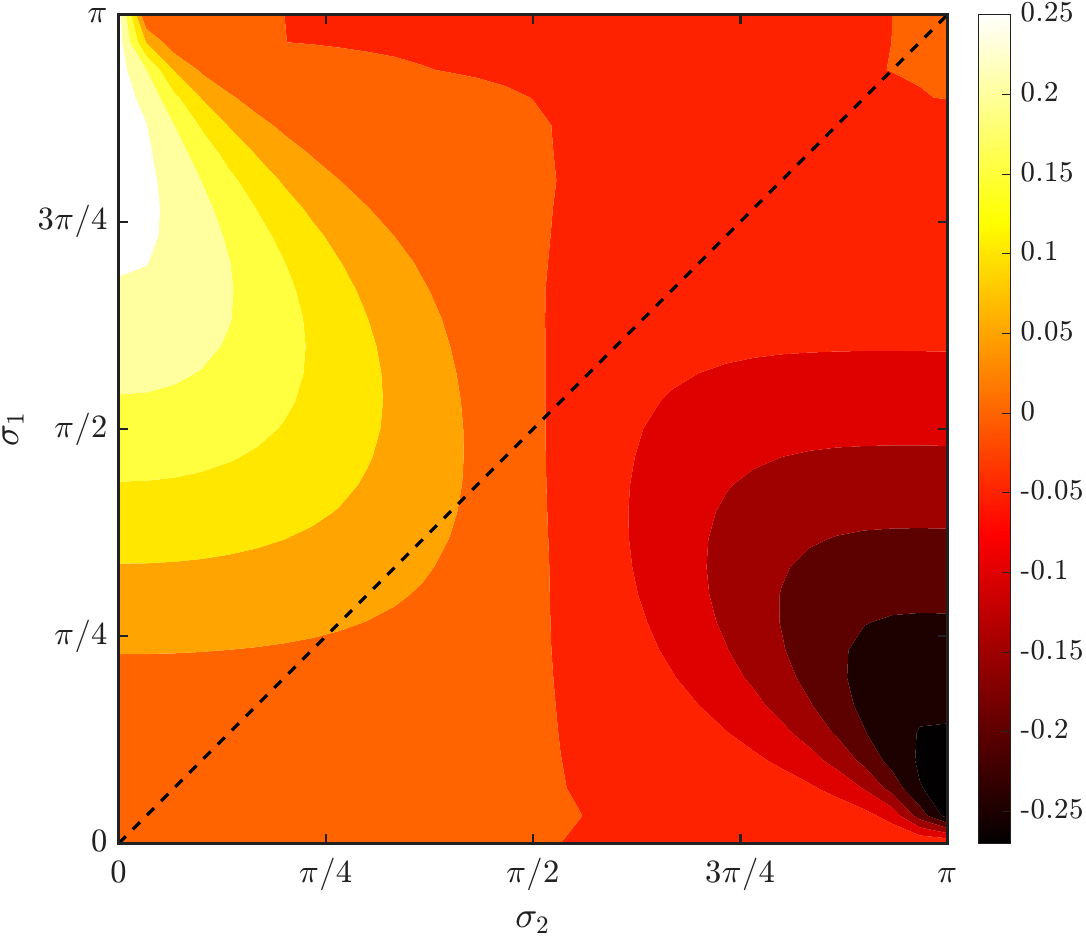}
    \caption{$(\kappa_1,\kappa_2) =(0.5,3)$}
    \label{fig:contour_sigma_2}
\end{subfigure}
&
\begin{subfigure}{0.33\textwidth}
    \centering
    \includegraphics[clip,width=\textwidth]{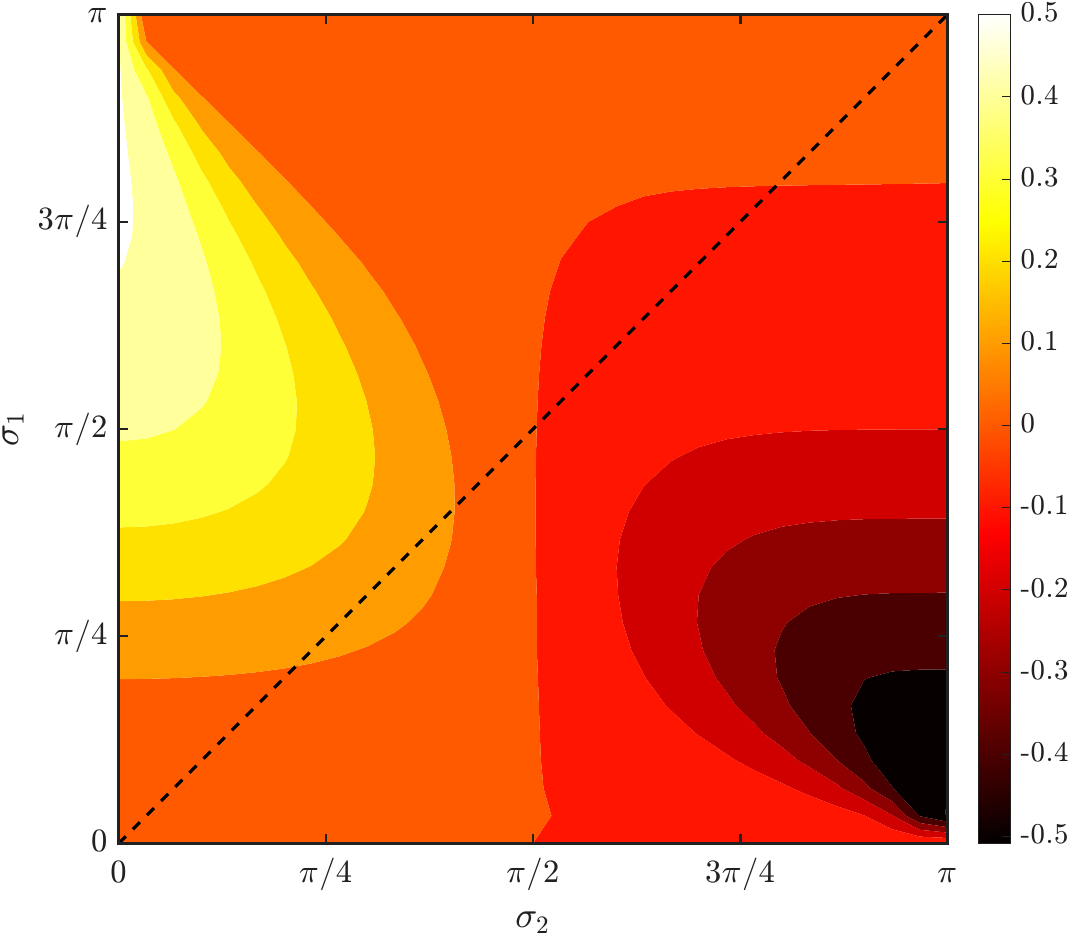}
    \caption{$(\kappa_1,\kappa_2) =(0.5,6)$}
    \label{fig:contour_sigma_3}
\end{subfigure}
\end{tabular}

\caption{ Contour plots of the time-averaged subsystem entropy difference $\Delta S$ as a function of the beam splitter angles $\sigma_1$ and $\sigma_2$. The factors $\cos^2(\sigma_1/2)$ and $\sin^2(\sigma_1/2)$ determine the weights of the evolutions governed by the chaos parameters $\kappa_1$ and $\kappa_2$, respectively; thus, moving along the $y$-axis increases the relative contribution of the $\kappa_2$ evolution. Similarly, $\cos^2(\sigma_2/2)$ and  $\sin^2(\sigma_2/2)$ control the post-selection weights associated with $\kappa_1$ and $\kappa_2$, respectively, so that increasing $\sigma_2$ along the $x$-axis favors the $\kappa_2$ evolution. The spin is initialized at $(\theta,\phi) = (2.25,1.1)$,  with spin size $j=25$, and the time average is performed over 200 kicks. The black dotted line corresponds to the symmetric case $\sigma_1=\sigma_2=\sigma$ discussed earlier.}
\label{fig:contour_plot}
\end{figure*}

After obtaining the spin density matrices for the superposition and classical--mixture cases, we compute the time-averaged subsystem entropy difference $\Delta S$, averaged over 200 time periods, for every combination of $\sigma_{1,2}$. This analysis is repeated for different $\kappa$ pairs, keeping $\kappa_1 = 0.5$ fixed in the regular regime and varying $\kappa_2$, as shown in Fig.~\ref{fig:contour_plot}. The black dotted line corresponds to the previously studied case $\sigma_1 = \sigma_2$.

Fig.~\ref{fig:contour_sigma_1} shows the time--averaged entropy difference $\Delta S$ for $\kappa_1 = 0.5$ and $\kappa_2 = 1$. Although both $\kappa$ values lie in the classically regular regime, certain combinations of beam splitter angles yield negative values of $\Delta S$. The minimum value in this case is $-0.0141$, indicating that the subsystem entropy generated in the superposed evolution can exceed that of  the classical mixture. This occurs when the first beam splitter weights the smaller $\kappa$ branch more strongly ($\sigma_1$ small), whereas the post-selection amplifies the larger $\kappa$ value ($\sigma_2 \approx \pi$).

In Figs.~\ref{fig:contour_sigma_2} and \ref{fig:contour_sigma_3}, we repeat this analysis for $\kappa_2 = 3$ and $\kappa_2 = 6$, corresponding to classically mixed and fully chaotic phase space structures, respectively (see Figs.~\ref{fig:classical_3} and \ref{fig:classical_6}). In these cases, we observe a much larger negative entropy difference: down to $-0.2665$ for $\kappa_2 = 3$ and $-0.5020$ for $\kappa_2 = 6$. The region of negative $\Delta S$ in the $(\sigma_1,\sigma_2)$ plane is also significantly larger than in the $\kappa_2 = 1$ case. Notably, the mixed regime ($\kappa_2 = 3$) produces a broader area of negative $\Delta S$ than either the regular regime ($\kappa_2 = 1$) or the fully chaotic regime ($\kappa_2 = 6$), although the fully chaotic case yields the largest magnitude of negative entropy difference.


\section{Discussion}

\begin{table*}[t]
\centering
\renewcommand{\arraystretch}{1.25}
\setlength{\tabcolsep}{8pt}

\begin{tabular}{>{\raggedright\arraybackslash}p{0.16\textwidth}
                >{\raggedright\arraybackslash}p{0.14\textwidth}
                >{\raggedright\arraybackslash}p{0.18\textwidth}
                >{\raggedright\arraybackslash}p{0.40\textwidth}}
\toprule
\textbf{Chaos parameters} & \textbf{Beam splitter angles ($\sigma_1,\sigma_2$)} & \textbf{$\Delta S$} & \textbf{Physical interpretation} \\
\midrule

\multirow{7}{*}{$(\kappa_1,\kappa_2)=(0.5,\,1.0)$}
& $\sigma_1=\sigma_2$ 
& $\Delta S>0$
&  {Both arms exhibit predominantly regular dynamics; coherent superposition does not increase entropy production significantly relative to the classical mixture.} \\

\addlinespace[0.6ex]
\cmidrule(lr){2-4}
\addlinespace[0.6ex]

& $\sigma_1>\sigma_2$ 
& $\Delta S>0$, \quad ($\Delta S_{\max}\approx0.0434$) 
&  {Both arms exhibit predominantly regular dynamics; entropy generation in the case of superposition remains smaller but still comparable to that of the classical mixture}\\

\addlinespace[0.6ex]
\cmidrule(lr){2-4}
\addlinespace[0.6ex]

& $\sigma_1<\sigma_2$ 
& Mostly $\Delta S>0$ \quad ($\Delta S_{\min}\approx-0.0171$)
&   {Both arms exhibit predominantly regular dynamics; a small negative region appears only when post-selection favours the branch with larger $\kappa$ ($\sigma_2 \approx \pi$); the effect remains weak.} \\

\midrule

\multirow{7}{*}{$(\kappa_1,\kappa_2)=(0.5,\,3.0)$}
& $\sigma_1=\sigma_2$
& $\Delta S>0$ for $\sigma < \pi/2$;\; $\Delta S<0$ for $\sigma > \pi/2$
& {The $\kappa_2$ arm exhibits chaotic dynamics. When this branch dominates ($\sigma>\pi/2$), coherent superposition combined with post-selection increases entropy generation beyond that of the classical mixture.} \\

\addlinespace[0.6ex]
\cmidrule(lr){2-4}
\addlinespace[0.6ex]

& $\sigma_1>\sigma_2$
& $\Delta S>0$ for $\sigma_2<\pi/2$ \quad  $\Delta S<0$ for $\sigma_2>\pi/2$  \quad ($\Delta S_{\max} \approx 0.2713$)
&  {Mostly positive; negative values occur only when post-selection amplifies the chaotic branch ($\sigma_2 > \pi/2$); otherwise, the entropy generation remains comparable to the mixture case.} \\

\addlinespace[0.6ex]
\cmidrule(lr){2-4}
\addlinespace[0.6ex]

& $\sigma_1<\sigma_2$ 
& $\Delta S<0$ for $\sigma_2>\pi/2$ \quad  $\Delta S>0$ for $\sigma_2<\pi/2$  \quad ($\Delta S_{\min}\approx-0.2701$)
& {Mostly negative; when post-selection amplifies the chaotic branch ($\sigma_2 > \pi/2$), entropy production is higher for the case of coherent superposition.} \\

\midrule

\multirow{7}{*}{$(\kappa_1,\kappa_2)=(0.5,\,6)$}
& $\sigma_1=\sigma_2$
& $\Delta S>0$ for $\sigma<\pi/2$;\; $\Delta S<0$ for $\sigma>\pi/2$
&  {Classical mixture generates more subsystem entropy when regular evolution is favoured ($\sigma <\pi/2)$, while superposition increases it when the chaotic branch dominates.}{}\\

\addlinespace[0.6ex]
\cmidrule(lr){2-4}
\addlinespace[0.6ex]

& $\sigma_1>\sigma_2$
& Mostly $\Delta S>  0$, \quad ($\Delta S_{\max}\approx 0.5126$)
&  {Mostly positive; when post-selection amplifies the regular branch ($\sigma_2 < \pi/2$), entropy generation is maximally increased in the classical mixture compared to the superposed case.}\\

\addlinespace[0.6ex]
\cmidrule(lr){2-4}
\addlinespace[0.6ex]

& $\sigma_1<\sigma_2$ 
& Mostly $\Delta S <  0$, \quad ($\Delta S_{\min}\approx -0.5084$)
&  {Mostly negative; when post-selection amplifies the chaotic branch ($\sigma_2 > \pi/2$), entropy generation is maximally increased.}\\
\bottomrule
\end{tabular}

\caption{Summary of the sign of the single-qubit entropy difference $\Delta S$, its maximal and minimal values $\Delta S_{\max}$ and $\Delta S_{\min}$, for selected chaos parameter pairs ($\kappa_1,\kappa_2$) and beam splitter angles ($\sigma_1,\sigma_2$). Negative $\Delta S$ indicates increased entropy generation in the coherent superposition of evolutions compared to the classical mixture. Results are indicated for the initial state $(\theta,\phi)=(2.25,1.1)$, spin size $j=25$, and $200$ kicks (cf.~Fig.~\ref{fig:contour_plot}).
}
\label{tab:deltaS_summary_scan}
\end{table*}

We now turn to an analysis of  subsystem entropy generation in both the classical mixture and the quantum superposition  as functions of the beam splitter angles $\sigma_{1,2}$ and the difference in the chaos parameters $\kappa_2 - \kappa_1$. Table~\ref{tab:deltaS_summary_scan} clarifies the dynamical mechanism  underlying entropy generation in the interferometric protocol. The inequalities summarized in the table are not strict and are obtained from numerical analysis observed in the previous section.

\subsection{Same beam splitter angles ($\sigma_1 = \sigma_2 = \sigma$)}

Let us first analyze the case in which both beam splitter angles are equal,  $\sigma_1 = \sigma_2 = \sigma$. The density matrix for the classical mixture depends only on the first beam splitter angle, $\sigma=\sigma_1$, and consists of a mixture of two separately evolved density matrices under the kicked top unitary, weighted by $\cos^2(\sigma/2)$ and $\sin^2(\sigma/2)$, as given in Eq.~\eqref{eq: classical spin density}. Using the subadditivity of the von Neumann entropy, we can get the subsystem entropy as
\begin{equation}
    S(\rho^{cl}_{s_1}) \geq \cos^2{(\sigma_1/2)}S(\rho^{11}_{s_1}) + \sin^2{(\sigma_1/2)}S(\rho^{22}_{s_1})
\end{equation}
where $\rho^{cl}_{s_1}, \rho^{11}_{s_1}$ and $\rho^{22}_{s_1}$ are single qubit density matrices obtained after tracing out $(2j-1)$ qubits. As we increase the beam splitter angle $\sigma$, the lower bound on the subsystem entropy increases.

Next, we consider the coherently superposed evolutions followed by post-selection. In this case, the spin density matrix contains two diagonal terms, $\rho^{11}_s$ and $\rho^{22}_s$, whose prefactors scale as $\cos^4(\sigma/2)$ and $\sin^4(\sigma/2)$, respectively, while the interference term scales as $\sin^2(\sigma)$, as shown in Eq.~\eqref{eq: normalized post selected spin density for same sigma}. The interference contribution is maximal at $\sigma = \pi/2$. The subsystem entropy difference $\Delta S$ vanishes for $\sigma = 0,\ \pi$ or when $\kappa_1 = \kappa_2$. Any deviation in $\kappa$ or $\sigma$ results in a nonzero $\Delta S$.

To examine the effect of the chaos parameter difference $\kappa_2 - \kappa_1$, we fix $\sigma = \pi/2$ and $\kappa_1=0.5$, and vary $\kappa_2$. In the classical mixture, the subsystem entropy then depends only on the dynamical regimes determined by the $\kappa$ values: it is small when both 
parameters lie in the regular regime, large when both are chaotic, and intermediate when they belong to different regimes.

In the superposition case, keeping the beam splitter angle at $\sigma=\pi/2$ and setting $\kappa_2-\kappa_1 >0 $ activates the interference term, which is maximal at $\sigma = \pi/2$, thereby contributing significantly to the subsystem entropy. However, for $\sigma=\pi /2$, due to the fourth-power dependence of the prefactors of the diagonal terms in Eq.~\eqref{eq: normalized post selected spin density for same sigma}, their weights are reduced to one quarter (in contrast to one half in the classical mixture). After tracing out $(2j-1)$ qubits, the single qubit density matrix has larger contribution from regular evolution when compared to the chaotic case. Consequently, the subsystem entropy in the superposition case is suppressed, leading to a peak in the subsystem entropy difference curve (see Fig.~\ref{fig:entropy_diff_vs_kappa_sigma_1}(a)).
As $\kappa_2$ increases into the chaotic regime, the subsystem entropy of the classical mixture grows due to the contribution of the chaotic density matrix $\rho^{22}_s$. 
A similar trend occurs in the superposition case. While the contribution of the chaotic component is reduced due to the fourth-power prefactor, the interference term continues to increase the entropy.
Overall, this results in a reduction of $\Delta S$. Nevertheless, numerical analysis shows that $\Delta S$ remains positive for all values of $\kappa_2$. As summarized in Table~\ref{tab:deltaS_summary_scan}, for $\sigma = \pi/2$, the subsystem entropy difference $\Delta S$ is positive for all pairs of chaos parameters, irrespective of whether they lie in the regular or chaotic regime.

To examine the effect of the beam splitter angle $\sigma$, we fix the difference in the chaos parameters $\kappa_2 - \kappa_1$ and vary $\sigma$ over the interval $[0,\pi]$. When both $\kappa$ values lie in the regular regime, the interference term has only a minor effect, and 
the subsystem entropy of the classical mixture remains larger than that of the superposition for all $\sigma$.

When one of the $\kappa$ values enters the mixed phase space regime ($\kappa = 3$) or the fully chaotic regime ($\kappa =6$), one branch exhibits chaotic behaviour, and the interference term becomes increasingly relevant, with its magnitude controlled by $\sigma$. As $\sigma$ increases, the weight of the chaotic branch in the classical mixture grows as $\sin^2(\sigma/2)$, leading to increased entropy production (see Eq.~\eqref{eq: classical spin density}). In contrast, in the superposition case, the contribution of $\rho^{22}_s$ is suppressed due to the fourth-power prefactor, while the interference term increases only as $\sin(\sigma)$ (see Eq.~\eqref{eq: normalized post selected spin density for same sigma}). Although the subsystem entropy increases, it remains smaller than that of the classical mixture. This behaviour explains the left peak around $\sigma=\pi/3$ in Fig.~\ref{fig:entropy_diff_vs_kappa_sigma_1}(a) when one $\kappa$ lies in the mixed or chaotic regime.

For $\sigma > \pi/2$, the chaotic branch dominates. In the superposition case, as $\sigma$ increases, the combined effect of $\rho^{22}_s$ and the interference term eventually overcomes the slower growth associated with the fourth-power prefactor, in contrast to the quadratic dependence in the classical mixture (see Eq.~\eqref{eq: classical spin density}). This leads to the dip observed around $\sigma= 3\pi/5$ in Fig.~\ref{fig:entropy_diff_vs_kappa_sigma_1}(a). Nevertheless, the increase in entropy remains smaller than in the classical mixture, resulting in the asymmetric shape of the curves shown in  Fig.~\ref{fig:entropy_diff_vs_kappa_sigma_1}(a). Physically, increasing $\sigma$ amplifies the contribution of the chaotic branch, and post-selection further favours this branch. Consequently, the subsystem entropy in the superposition case increases, although it remains below that of the classical mixture. We therefore conclude that, for symmetric beam splitter angles ($\sigma_1=\sigma_2$), the superposition of a regular and a chaotic evolution yields increased subsystem entropy relative to the classical mixture only when the chaotic branch carries the larger weight and is reinforced by post-selection ($\sigma>\pi/2)$.

\subsection{Different beam splitter angles ($\sigma_1 \neq \sigma_2$)}

We now consider the asymmetric case, $\sigma_1 \neq \sigma_2$. In this regime, the subsystem entropy difference $\Delta S$ exhibits nontrivial structure over the $(\sigma_1,\sigma_2)$ plane. The classical mixture is independent of the second beam splitter angle $\sigma_2$, whereas in the superposition followed by post-selection the interference term depends on both beam splitter angles through a sine modulation. 

When both chaos parameters lie in the regular regime, $(\kappa_1,\kappa_2) = (0.5,1)$, almost the entire plane is characterized by positive $\Delta S$, with only a small region of negative $\Delta S$ for $\sigma_1 < \pi/3$ and $\sigma_2 \approx \pi$ (see Fig.~\ref{fig:contour_sigma_1}). This behaviour can be understood similarly to the symmetric case: since both $\kappa$ values correspond to regular dynamics, the classical mixture generates more subsystem entropy than the superposition with post-selection.

When $\kappa_2$ enters the mixed or chaotic regime, the contour structure of $\Delta S$ develops pronounced positive and negative islands. As shown in Figs.~\ref{fig:contour_sigma_2} and \ref{fig:contour_sigma_3} for $(\kappa_1,\kappa_2) = (0.5,3)$ and $(0.5,5)$, respectively, extended regions of alternating sign appear. To understand these structures, we fix $\sigma_1$ and vary $\sigma_2$. Since the classical mixture does not depend on $\sigma_2$, its subsystem entropy remains constant along horizontal cuts in the contour plot. In contrast, in the superposition case, $\sigma_2$ modifies the post-selection and hence the relative weight of regular and chaotic dynamics.

Consider, for example, a horizontal cut at $\sigma_1 = \pi/8$ in Figs.~\ref{fig:contour_sigma_2} and \ref{fig:contour_sigma_3}. As $\sigma_2$ is varied, post-selection favours regular dynamics for $\sigma_2 < \pi/2$ and chaotic dynamics for $\sigma_2 > \pi/2$. The interference term is maximal at $\sigma_2 = \pi/2$. For $\sigma_2 = \pi$, however, the interference term vanishes and the normalized spin density matrix reduces to $\rho^{22}_s$, corresponding to purely chaotic evolution (see Eq.~\eqref{eq: normalized post selected spin density}). In this regime, the superposition with post-selection generates more entropy than the classical mixture, leading to $\Delta S < 0$, with larger magnitude for higher $\kappa_2$ values (e.g., $\kappa_2 = 6$).

Similarly, fixing $\sigma_1 = 3\pi/4$ and varying $\sigma_2$, we observe that $\Delta S$ decreases as post-selection increasingly favours the chaotic branch. The maximum positive $\Delta S$ occurs for $\sigma_2 = 0$ together with $\sigma_1 = 3\pi/4$. In this configuration, the interference term vanishes and the post-selection effectively projects onto $\rho^{11}_s$, which evolves under the regular parameter $\kappa_1 = 0.5$ (see Eq.~\eqref{eq: normalized post selected spin density}). The classical mixture, in contrast, still contains contributions from both regular and chaotic evolutions.

An important intuition emerges from this asymmetric configuration. Choosing $\sigma_1$ small biases the initial splitting toward the regular branch, so that the evolution is predominantly regular for most of the protocol. However, selecting a large value of $\sigma_2$ at the final stage ($\sigma_2 \approx \pi$) causes the post-selection to favour the chaotic branch. Consequently, the final spin state can display increased entropy production characteristic of chaotic dynamics, even though the evolution was initially weighted toward regular behaviour.

In this sense, the second beam splitter does not merely reveal pre-existing dynamics but reshapes the effective contribution of the two branches through coherent recombination and post-selection. This mechanism is responsible for the pronounced negative regions of $\Delta S$ observed in the contour plots.

This behaviour is closely related to the \textit{quantum eraser} effect \cite{scully_quantum_erasure_1982}. In a double-slit experiment, acquiring which-path information suppresses interference, whereas erasing that information restores it. In our interferometric protocol, the two kicked top evolutions play the role of the two paths, and the second beam splitter together with post-selection determines whether the dynamical branches are coherently recombined or effectively distinguished.

When $\sigma_2 = 0$ or $\pi$, the interference term in Eq.~\eqref{eq: normalized post selected spin density} vanishes, and the final spin state reduces to a single dynamical evolution. This limit is analogous to obtaining which-path information. For intermediate values of $\sigma_2$, the interference term survives, and coherent recombination modifies the subsystem entropy. In this way, varying $\sigma_2$ controls the extent to which dynamical \enquote{which-branch} information is erased, thereby tuning the interference contribution to entropy generation.

\section{Summary}

In previous studies, the Mach–Zehnder interferometer has been primarily employed for high-precision phase estimation, often incorporating noise models or analyzing the impact of different input states on metrological performance~\cite{MZI_dowling_2000,MZI_phase_bouwmeester_2007,MZI_phase_daniel_2013}. 
In contrast, we use the Mach–Zehnder setup to investigate the effect of \emph{superposing two distinct quantum evolutions}. Instead of introducing a relative phase shift between the arms, we insert two quantum kicked top unitaries with different Hamiltonian parameters into the two paths of the interferometer. Since the classical kicked top exhibits qualitatively different dynamics -- regular, mixed, or chaotic -- depending on the kicking strength, this setup allows us to probe how classical dynamical properties influence the resulting superposed quantum evolution. To the best of our knowledge, this is the first study to systematically explore the superposition of regular and chaotic quantum evolutions within such an interferometric framework.

We use the time-averaged subsystem entropy (see Eq.~\eqref{eq:single qubit entropy difference}) as a probe to quantify the difference between the superposition of two evolutions and their classical mixture. Our results indicate that when both arms of the interferometer remain in the regular regime, coherent superposition does not provide an advantage over the classical mixture, and time-averaged subsystem entropy $\Delta S$ remains positive. 
Once one branch enters the chaotic regime, a crossover emerges: if the chaotic evolution carries sufficient weight in the interferometer, the coherent superposition generates more subsystem entropy than the classical mixture, resulting in $\Delta S < 0$. The effect becomes more pronounced as the degree of chaos increases (e.g., $\kappa_2 = 6$), indicating that quantum interference between regular and chaotic dynamics increases entropy production. By contrast, when both evolutions belong to the same dynamical class -- either both regular or both chaotic -- the superposition produces \emph{less} subsystem entropy than classical mixing. This indicates that the interference term in the post-selected spin state contributes significantly to entropy production primarily when the two evolutions differ in their underlying classical dynamics. This trend is clearly reflected in Figs.~\ref{fig:entropy_diff_vs_kappa_sigma_1} and \ref{fig:entropy_diff_vs_kappa_sigma_2}, and in scans across different classical phase space structures (see Fig.~\ref{fig:entropy_diff_vs_phi}).

For the asymmetric case ($\sigma_1\neq \sigma_2$), the classical mixture generally generates more subsystem entropy than the superposition with post-selection, except within a limited parameter region that warrants 
further investigation. When one of the two chaos parameters lies in the mixed or chaotic regime, the superposition can generate more subsystem entropy, provided that post-selection strongly favours the branch with the larger $\kappa$. A representative example occurs in the mixed and fully chaotic cases ($\kappa_2=3,\ 6$), where the maximal subsystem entropy in the superposed scenario appears near $\sigma_1\approx \pi/8$ and $\sigma_2\approx\pi$ (see Figs.~\ref{fig:contour_sigma_2} and \ref{fig:contour_sigma_3}). This suggests that optimal entropy generation through superposition is achieved by initially weighting the state toward the regular evolution while post-selecting onto the chaotic branch. As discussed earlier, this behaviour can be interpreted in terms of the quantum eraser effect \cite{scully_quantum_erasure_1982}. By suitably choosing the post-selection, one can effectively increase or suppress the chaotic contribution in the superposed evolution, analogous to some form of \textit{chaos erasure} in the spin system.

Overall, our results reveal a direct connection between dynamical instability and quantum coherence, as summarized in Table~\ref{tab:deltaS_summary_scan}. Even for modest spin sizes, well away from the semiclassical limit, the interference between quantum evolutions remains strongly influenced by the underlying classical dynamics. Some effects are expected to become sharper with increasing spin size, motivating a careful analysis of finite-size effects. An important open question is how sharply the subsystem-entropy transition develops in the large-spin (semiclassical) limit.

Several additional directions merit investigation. One avenue is to explore the role of the ancilla dimension by superposing multiple evolutions simultaneously, and to consider the continuous-variable ancilla limit in which the $\kappa$-space becomes effectively infinite dimensional. A more rigorous analytical treatment could be pursued using simplified toy models that allow for greater numerical control. Finally, it would be valuable to assess the feasibility of implementing related interference protocols in experimental platforms such as cold-atom or Bose–Einstein condensate systems.

\section{Acknowledgements}
ACdlH is supported by the Institute for Theoretical Studies at ETH Zurich through a Junior Research Fellowship.
This work was supported in part by the Natural Sciences and Engineering Research Council of Canada (NSERC).  Wilfrid Laurier University and the University of Waterloo are located in the traditional territory of the Neutral, Anishnawbe and Haudenosaunee peoples. We thank them for allowing us to conduct this research on their land. This research was funded in part by the Austrian Science Fund (FWF) [10.55776/F71] and [10.55776/COE1]. 



\balance

\nocite{apsrev42Control}
\bibliography{ref}

\clearpage
\onecolumngrid

\setlength{\textfloatsep}{8pt plus 2pt minus 2pt}
\setlength{\floatsep}{6pt plus 2pt minus 2pt}
\setlength{\intextsep}{6pt plus 2pt minus 2pt}

\section*{Supplementary Material}

\setcounter{figure}{0}
\renewcommand{\thefigure}{S\arabic{figure}}

\begin{figure}[!ht]
\centering
\setlength{\tabcolsep}{3pt}
\renewcommand{\arraystretch}{1.0}

\begin{tabular}{ccc}
\includegraphics[trim={100mm 8mm 100mm 13mm},clip,width=0.32\linewidth]{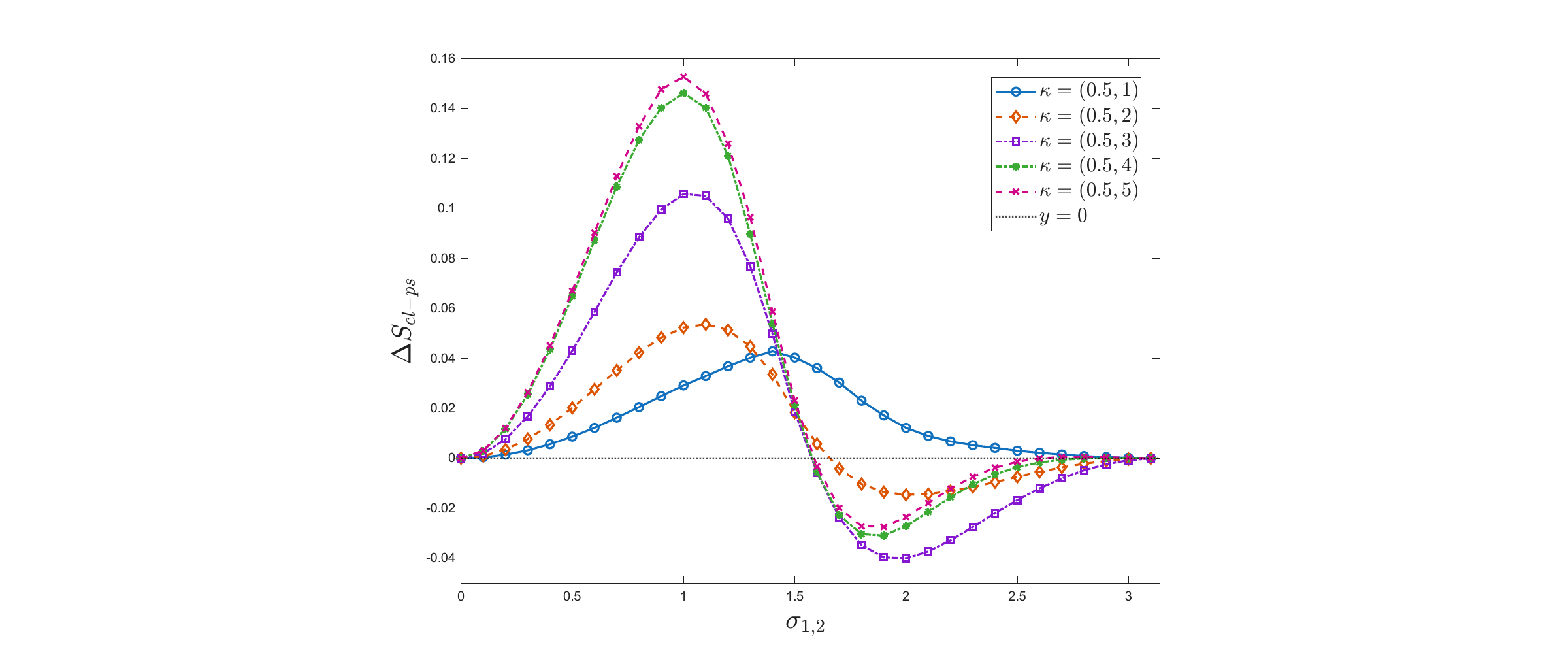} &
\includegraphics[trim={100mm 8mm 100mm 13mm},clip,width=0.32\linewidth]{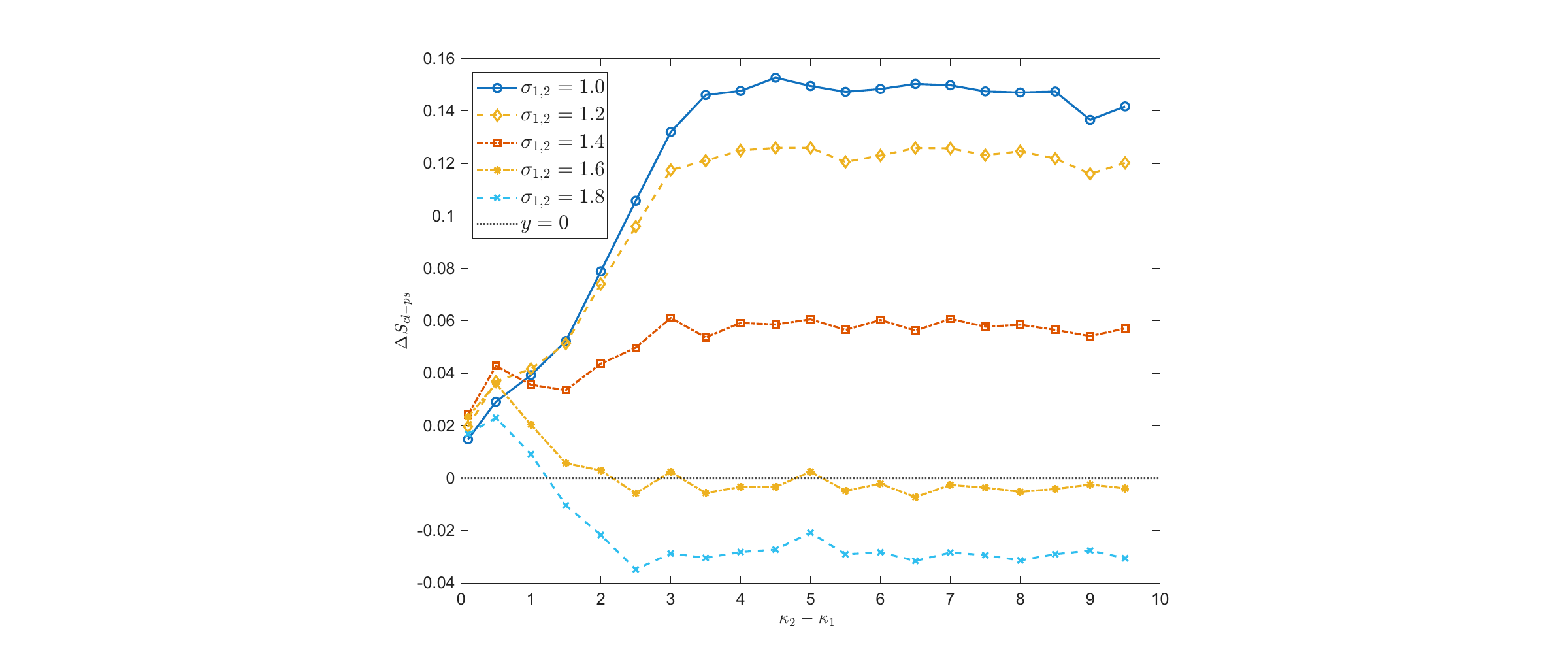} &
\includegraphics[trim={97mm 7mm 95mm 13mm},clip,width=0.33\linewidth]{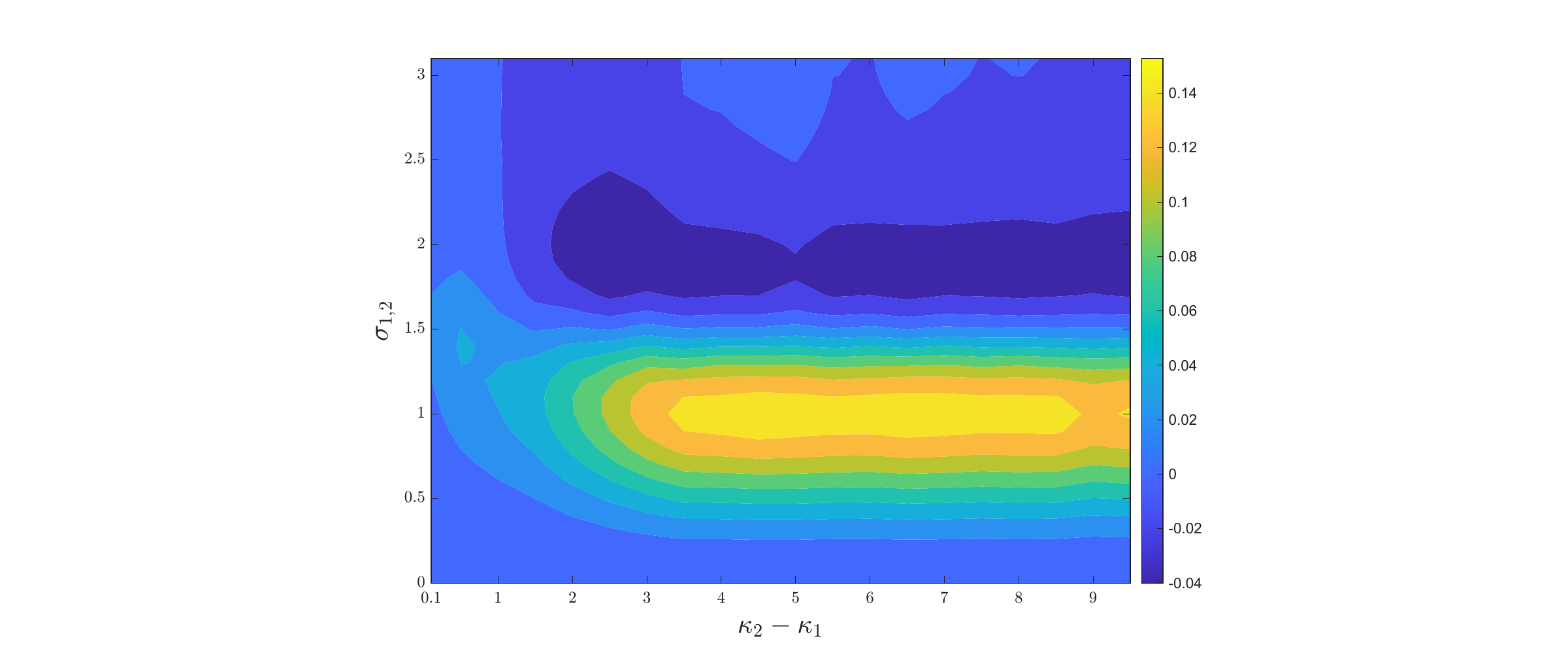} \\
(a) & (b) & (c)
\end{tabular}

\caption{
Time-averaged subsystem entropy difference $\Delta S$, averaged over 200 periods of the kicked top unitary with $j=25$. The spin coherent state is initialized at $(\theta,\phi) = (2.25, -1.6)$, and the ancilla qubit is initialized in the state $\ket{1}$. 
(a) $\Delta S$ versus beam splitter angles $\sigma_1=\sigma_2$ (31 samples in $(0,\pi)$). 
(b) $\Delta S$ versus $\kappa_2-\kappa_1$ (20 samples in $(0.1,9.5)$). 
(c) Contour plot in the $(\sigma,\kappa_2-\kappa_1)$ plane. 
The black dotted line indicates $\Delta S=0$ in panels (a) and (b).
}
\label{fig:entropy_diff_vs_kappa_sigma_3}
\end{figure}

Here we report results for a different initial spin coherent state $(\theta,\phi) = (2.25,-1.6)$, 
shown in Fig.~\ref{fig:entropy_diff_vs_kappa_sigma_3}. In the classical limit, the dynamics of this initial state also exhibits a regular-to-chaotic transition as $\kappa$ increases. The qualitative features of subsystem entropy difference $\Delta S$ are essentially the same as for the $(\theta,\phi) = (2.25,1.1)$ case as discussed in the main text.

\begin{figure}[!ht]
\centering
\setlength{\tabcolsep}{3pt}
\renewcommand{\arraystretch}{1.0}

\begin{tabular}{ccc}
\includegraphics[trim={100mm 8mm 100mm 13mm},clip,width=0.32\linewidth]{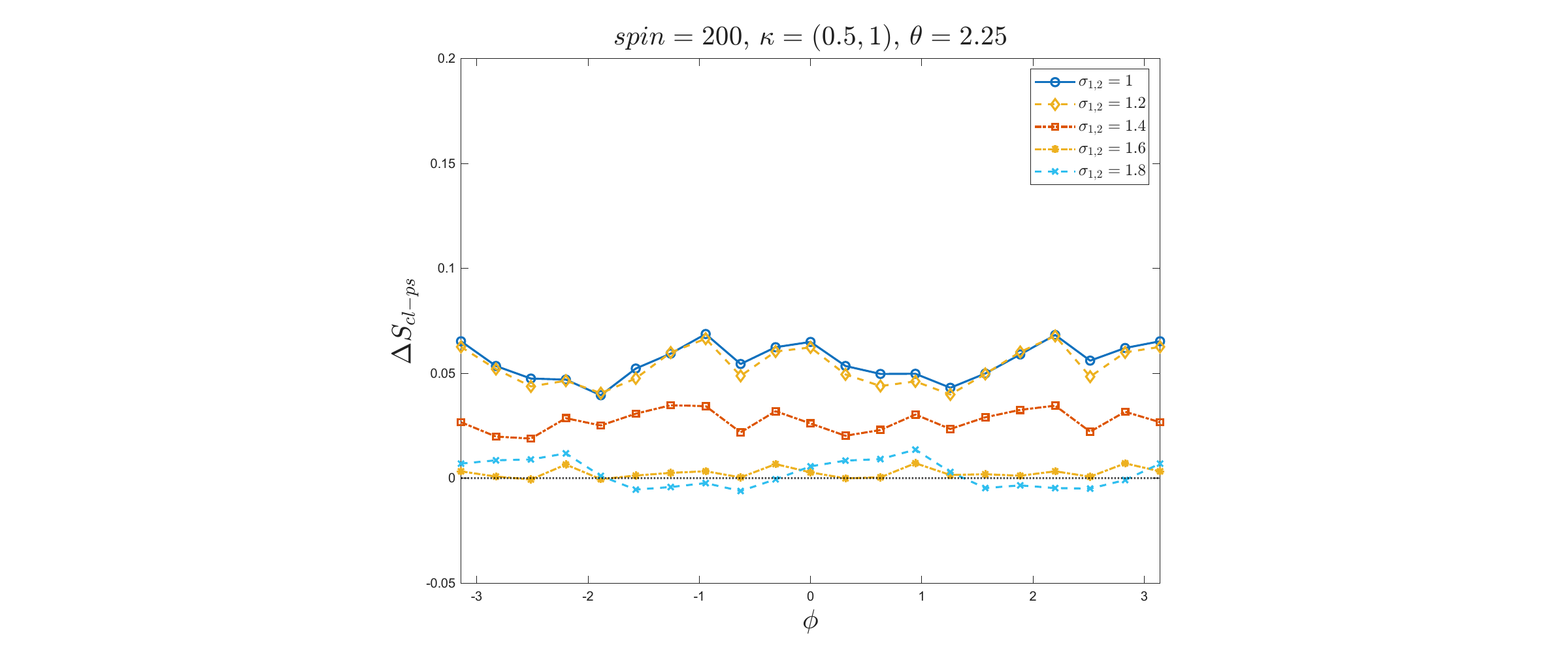} &
\includegraphics[trim={100mm 8mm 100mm 13mm},clip,width=0.32\linewidth]{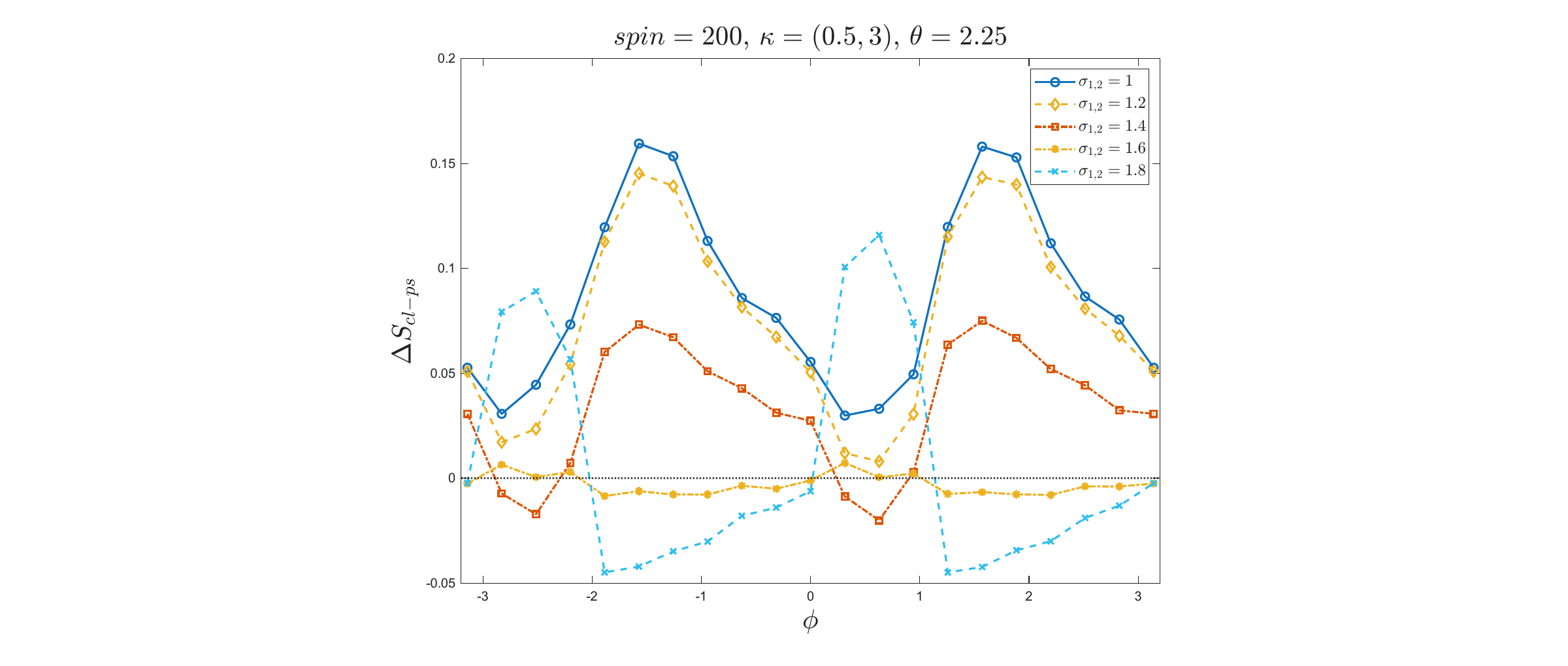}  &
\includegraphics[trim={97mm 7mm 95mm 13mm},clip,width=0.33\linewidth]{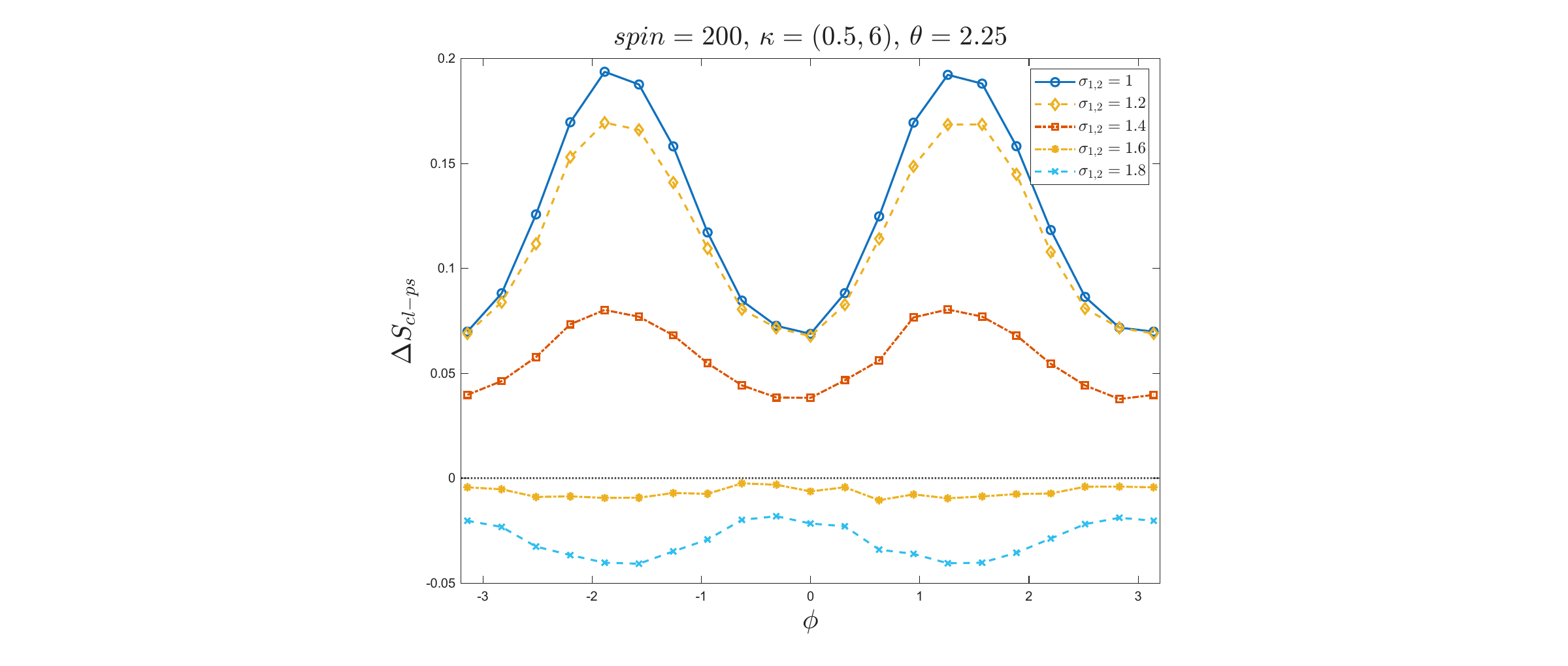}  \\
(a) & (b) & (c)
\end{tabular}

\caption{
Time-averaged subsystem entropy difference $\Delta S$, averaged over 200 periods of the kicked top unitary for $j=200$. 
Panels (a)–(c) correspond to $(\kappa_1,\kappa_2)=(0.5,1.0)$, $(0.5,3.0)$, and $(0.5,6.0)$, respectively. 
Different curves correspond to beam splitter angles with $\sigma_1=\sigma_2$. 
For each curve, 21 spin coherent states are initialized at $\theta=2.25$ with uniformly spaced $\phi\in[-\pi,\pi]$. 
The black dotted line indicates $\Delta S=0$.
}
\label{fig:entropy_diff_vs_phi_2}
\end{figure}

In Fig.~\ref{fig:entropy_diff_vs_phi_2} we depict results for spin size $j=200$. We see that the spurious finite-size effects we commented on in Sec.~\ref{sec: Different initial states, same sigma} are removed, as expected. We find, as before, that superposing chaotic and regular evolutions leads to greater subsystem entropy generation than the classical mixture, provided the chaotic contribution dominates.

\vspace{1cm}

\vspace{1cm}

\end{document}